\documentclass[11pt, reqno]{amsart}
\usepackage{amsfonts}
\usepackage{amssymb}
\usepackage{latexsym}
\usepackage[final]{hyperref}
\usepackage[]{enumerate}
\usepackage{verbatim}
\usepackage[english]{babel} 
\usepackage{blindtext}
\usepackage[nohyperlinks,nolist]{acronym}
\usepackage{listings}
\usepackage{multirow}
\usepackage{algorithm}
\usepackage{algpseudocode}

\algnewcommand\algorithmicparameters{\textbf{Parameters:}}
\algnewcommand\Parameters{\item[\algorithmicparameters]}

\usepackage[table]{xcolor}%

\usepackage{graphicx}       
\usepackage[utf8]{inputenc} 
\usepackage{amsmath}

\usepackage{tikz}
\usepackage{cmap}
\usepackage{ulem}
\usepackage{soul}
\usepackage{comment}
\usepackage{fancybox}
\usepackage{relsize}
\usepackage{pifont}
\usepackage{subcaption}

\newcommand{\norm}[1]{\left\Vert #1 \right\Vert}

\newcommand{\Innerprod}[2]{\left\langle \, #1 , #2 \, \right\rangle}

\DeclareMathOperator{\diag}{diag}

\DeclareMathOperator{\trace}{tr}

\DeclareMathOperator{\grad}{grad}
\DeclareMathOperator{\Proj}{Proj}

\DeclareMathOperator{\Prec}{Prec}
\DeclareMathOperator{\Sym}{Sym}
\DeclareMathOperator{\St}{St}
\DeclareMathOperator{\Gr}{Gr}
\DeclareMathOperator{\Spec}{Spec}
\DeclareMathOperator{\lift}{lift}

\theoremstyle{definition}

\theoremstyle{remark}

\usepackage{graphicx} 

\newcommand{\vampyr}{\texttt{VAMPyR}}

\newcommand{\mrchem}{\texttt{MRChem}}
\newcommand{\mrchemsoft}{\texttt{MRChemSoft}}

\begin{acronym}
\acro{CASSCF}{complete active space self-consistent field}
\acro{DFT}{density functional theory}
\acro{DIIS}{direct inversion in the iterative subspace}
\acro{GTO}{Gaussian Type Orbital}
\acro{KAIN}{Krylov-accelerated inexact Newton}
\acro{MRA}{multiresolution analysis}
\acro{NS}{non-standard}
\acro{ROHF}{restricted open-shell Hartree-Fock}
\acro{SAD}{superposition of atomic densities}
\acro{SCF}{self-consistent field}
\end{acronym}

\title
[Gradient descent for Hartree-Fock theory]
{Riemannian gradient descent for Hartree-Fock theory}

\author[Dinvay]{Evgueni Dinvay}

\email{ evgueni.dinvay@gmail.com }

\address
{
    Department of Chemistry
    \\
    UiT The Arctic University of Norway
    \\
    PO Box 6050 Langnes
    \\
    N-9037 Tromsø
    \\
    Norway
}


\begin{document}

\begin{abstract}
We present a Riemannian optimization framework for Hartree-Fock theory
formulated directly in the Sobolev space $H^1$.
The orthonormality constraints are interpreted geometrically
via infinite-dimensional Stiefel and Grassmann manifolds
endowed with the embedded $H^1$ metric.
Explicit expressions for Euclidean and Riemannian gradients,
tangent-space projections, and retractions are derived using resolvent operators,
avoiding distributional formulations.
The resulting algorithms include Riemannian steepest descent
and a preconditioned nonlinear conjugate gradient method equipped with
Armijo backtracking and Powell-type restarts.
Particular attention is given to physically motivated preconditioning
based on inversion of the kinetic energy operator.
The framework is naturally compatible with adaptive multiwavelet discretizations,
where Coulomb-type convolutions can be evaluated efficiently.
Numerical experiments demonstrate robust convergence
and competitive performance compared to conventional SCF-DIIS schemes.
In addition, for small molecules the gradient descent method converges
from random initial guesses.
The proposed formulation provides a geometrically consistent
and discretization-independent perspective on electronic structure optimization
and offers a foundation for further developments
in infinite-dimensional Riemannian methods for quantum chemistry.
\end{abstract}

\maketitle
\section{Introduction}
\setcounter{equation}{0}

\subsection{Variational formulation}

The variational approach in quantum chemistry
is a formulation of the ground state problem of a molecular system as minimization
of the Hamiltonian expectation value subject to normalization of the wave function.
In practice, it is simplified as the orbital optimization problem:
\begin{equation}
\label{orbital_optimization_problem}    
    \text{minimize }
    \,
    \mathcal E(\phi)
    \,
    \text{ with the orthonormal orbital vector }
    \,
    \phi
    =
    (\phi_1, \ldots, \phi_N)^T
    .
\end{equation}
In general,
the orbitals $\phi_j$ depend on three spatial coordinates and one spin coordinate,
making a total of four variables.
In this work we focus on a restricted setting of the spatial
orbitals $\phi_j(x_1, x_2, x_3)$ which are real-valued square-integrable functions.
Such simplification is considered to be a good model for closed-shell molecules.
The orthonormality constraint mentioned in \eqref{orbital_optimization_problem} reads as
\begin{equation}
\label{orthonormality_constraint}    
    \Innerprod{\phi_i}{\phi_j}_{L^2}
    =
    \int \phi_i \phi_j
    =
    \delta_{ij}
    , \quad
    \text{ for }
    \,
    i, j = 1, \ldots, N.
\end{equation}

A concrete form of the energy functional $\mathcal E$ depends on the approximation paradigm.
The vast majority of functionals,
including the restricted Hartree-Fock energy
\begin{equation}
\label{hartree_fock_energy}
    \mathcal E(\phi)
    =
    2
    \sum_{i = 1}^N
    (i |h| i)
    +
    \sum_{i, j = 1}^N
    (
        2
        (ii|jj)
        -
        (ij|ij)
    )
    ,
\end{equation}
are well defined for smooth functions $\phi_j$.
Here in \eqref{hartree_fock_energy}
we use common chemical notation \cite{Szabo_Ostlund}
for one and two body integrals.
They will be recalled below in the text,
when we differentiate the expression \eqref{hartree_fock_energy}.
Among different spaces of orbitals the crucial place is taken
by the product of Sobolev spaces $H^1 \left( \mathbb R^3 \right)$,
defined by the inner product
\begin{equation}
\label{Sobolev_inner_product}
    \Innerprod{\varphi}{\psi}_{H^1}
    =
    \Innerprod{\nabla \varphi}{\nabla \psi}_{L^2}
    +
    \Innerprod{\varphi}{\psi}_{L^2}
    ,
\end{equation}
turning the entire product
\(
    H^1 \left( \mathbb R^3 \right)
    \times \ldots \times
    H^1 \left( \mathbb R^3 \right)
\)
into a Hilbert space that we simply
denote as $H^1$.
In physical terms,
this corresponds to orbitals with square-integrable gradients --- exactly the level of regularity required for kinetic energy to be well-defined.
Sobolev orbitals are orbitals with finite kinetic energy.
Importantly,
the variational problem has minimizers in the Sobolev space $H^1$
for the majority of physically relevant energy functionals $\mathcal E(\phi)$,
including Hartree-Fock \cite{Lions1987},
Kohn-Sham \cite{Anantharaman_Cances2009} and multiconfigurational \cite{Friesecke2003} models.
Moreover,
these functionals are differentiable with respect to $H^1$-norm.

Throughout the text the Hartree-Fock functional \eqref{hartree_fock_energy}
will serve as an illustrative example.
Nevertheless,
all the theory extends to the Kohn-Sham functionals.
In particular,
numerical simulations with B3LYP will be conducted below.
The number of electrons is $2N$.
Multiconfigurational models are not considered in this paper.

The orthonormality constraint \eqref{orthonormality_constraint},
imposed on the orbital vector
\(
    \phi
    =
    (
        \phi_1
        , \ldots ,
        \phi_N
    )^T
\)
in \eqref{orbital_optimization_problem},
forms a surface in a space of $N$-vector functions,
which we refer to as a Stiefel manifold.
In the case of a one orbital model, associated with two electron systems,
the Stiefel manifold coincides with the unit sphere,
corresponding to the constraint $\norm{\phi}_{L^2} = 1$.
For some numerical algorithms solving
the variational problem \eqref{orbital_optimization_problem},
we need to specify geometry of the orthonormality constraint surface,
which transforms the surface into a rich geometrical object,
known as a Riemannian manifold \cite{Boumal2023}.
The choice may affect the computational efficiency.
The most natural geometry for the $L^2$-orthogonality \eqref{orthonormality_constraint}
is $L^2$-based metric,
which is mostly used in practice.
On the other hand,
the energy functional $\mathcal E(\phi)$ is not differentiable with respect to $L^2$-norm,
which suggests to embed the Stiefel manifold in $H^1$ space instead.
Then both the functional and the constraint surface of
the variational problem \eqref{orbital_optimization_problem} are smooth,
which is a completely unexplored perspective.
This view could significantly enrich several methods based on different
discretization techniques, especially
the ones imitating {\it the complete basis set limit}:
multiwavelets \cite{Bischoff2019, Harrison_Fann_Yanai_Gan_Beylkin2004}
and finite elements \cite{Kuang_Shen_Hu2024, Motamarri_Nowak_Leiter_Knap_Gavini2013}.

The scope of the current research is the continuous functional view
of the optimization problem \eqref{orbital_optimization_problem}.
It is worth, however, to draw a connection to the common practice
in computational chemistry.
By reducing the search space,
or more precisely the domain of $\mathcal E(\phi)$,
to a finite dimensional function space one arrives
at a basis-discretized variational formulation \cite{Helgaker_Jorgensen_Olsen}.
In an orthonormal basis of size $N_b$
a variable $\phi$ can be viewed as an $N_b \times N$-matrix,
often referred to as a molecular orbital.
From this perspective
the Stiefel manifold is exactly the $N_b \times N$-matrices $\phi$
satisfying $\phi^T \phi = 1 (\text{identity})$.
In other words,
the molecular orbital formalism is the minimization problem
constrained to the finite dimensional Stiefel manifold.
It turns out that further changing variables, one can completely eliminate
the constraint.
Indeed,
fixing an arbitrary molecular orbital $\phi^{(0)}$ one
can parametrize the whole Stiefel manifold by the linear space
of skew symmetric matrices $\kappa = - \kappa^T$ as
\(
    \phi = e^{- \kappa} \phi^{(0)}
    .
\)
This approach is called orbital rotation representation.

\subsection{Optimization methods}

The dominant difficulty of the optimization problem
\eqref{orbital_optimization_problem} is that the search space is a nonlinear surface.
Moreover, the orbital dependence of the energy functional
is physically meaningful only on the orthogonality constraint surface;
its extension outside this manifold bears no relation to the original Hamiltonian.
All numerical algorithms for the optimization problem
\eqref{orbital_optimization_problem} are grounded in two
fundamentally different methods: fixed-point iteration and gradient descent.
The former
accompanies
the Euler-Lagrange formalism, which recasts
the constraint optimization \eqref{orbital_optimization_problem}
into searching for stationary points
\begin{equation}
\label{Euler_Lagrange_formulation}
    \nabla \mathcal L(\phi, \varepsilon) = 0
    \,
    \text{ with the Lagrangian functional }
    \,
    \mathcal L(\phi, \varepsilon)
    =
    \mathcal E(\phi)
    -
    \sum_{i,j} \varepsilon_{ij} \bigl( \Innerprod{\phi_i}{\phi_j} - \delta_{ij} \bigr)
    .
\end{equation}
A naive evaluation of the gradient with respect to $L^2$-inner product
leads to the following nonlinear eigenvalue problem
\begin{equation}
\label{nonlinear_eigenvalue_problem}
    F(\phi) \phi = \varepsilon \phi
    .
\end{equation}
In the functional formalism
\(
    F(\phi) = - \Delta / 2 + V(\phi)
\)
and
Equation \eqref{nonlinear_eigenvalue_problem} can only be understood
in the generalized sense,
or as an equation in the negative order Sobolev space $H^{-1}$,
recall that the energy functional is neither differentiable
nor defined on $L^2$-functions.
The $H^s$-inner product with an $s \in \mathbb R$ is defined as an extension
of \eqref{Sobolev_inner_product} for $s = 1$ through the Fourier transform.
In the molecular orbital formalism \eqref{nonlinear_eigenvalue_problem}
is solved iteratively as
\begin{equation}
\label{SCF_procedure}    
    F \left( \phi^{(n)} \right) \phi^{(n+1)} = \varepsilon^{(n+1)} \phi^{(n+1)}
    ,
\end{equation}
which is known as a
\ac{SCF}
procedure.
As a matter of fact,
a precise definition of \ac{SCF} scheme depends on the choice of discretization.
In contrast to diagonalization of the Fock matrix \eqref{SCF_procedure}
in the molecular orbital representation,
in the multiwavelet framework \cite{Harrison_Fann_Yanai_Gan_Beylkin2004}
it is more feasible to precondition out
the kinetic energy operator in Equation \eqref{nonlinear_eigenvalue_problem}
in the following way
\begin{equation}
\label{preconditioned_nonlinear_eigenvalue_problem}
    \phi
    =
    - ( - \Delta / 2 - \varepsilon )^{-1} V(\phi)
\end{equation}
and run the fixed point iteration algorithm
\begin{equation}
\label{preconditioned_SCF_procedure}
    \phi^{(n+1)}
    =
    -
    \left(
        - \frac{ \Delta }2 - \varepsilon \left( \phi^{(n)} \right)
    \right)^{-1}
    V
    \left( \phi^{(n)} \right)
    , \quad
    \varepsilon_{ij} \left( \phi \right)
    =
    \Innerprod{(F(\phi) \phi)_i}{\phi_j}_{L^2}
    .
\end{equation}
Interestingly, Equation \eqref{preconditioned_nonlinear_eigenvalue_problem}
can be understood classically in the Sobolev space $H^1$.
Therefore, the Hartree-Fock method is intrinsically an $H^1$ orbital theory.
In theory, the algorithm \eqref{preconditioned_SCF_procedure} is applicable
regardless of the discretization method in use,
and could be considered as a general \ac{SCF} procedure.
In practice, \eqref{SCF_procedure} is more efficient
than \eqref{preconditioned_SCF_procedure}
in the case of \ac{GTO} discretization, for example.
Mathematically speaking, the \ac{SCF} procedure
is not a purely fixed point algorithm.
Indeed, in the form \eqref{preconditioned_SCF_procedure}
the main part of Hessian, the kinetic operator, is preconditioned out,
whereas in the form \eqref{SCF_procedure}
the whole Fock matrix is preconditioned out.
From this perspective,
\ac{SCF} should be viewed as a quasi-Newton method by default.
Furthermore,
in practical calculations \ac{SCF} is accelerated by
a particular linear extrapolation technique \cite{Pulay1980, Pulay1982}
called a \ac{DIIS},
by weighting the previous iterations according to their efficiency.
\ac{DIIS} is closely related to Anderson acceleration of fixed-point maps \cite{Anderson1965}.
As pointed out in \cite{Rohwedder_Schneider2011}
this linear extrapolation is similar to a projected quasi-Newton method.

The second foundational optimization method
and the main focus of the current work
is the gradient descent,
which in its simplest form can be formulated in the following way
\begin{equation}
\label{unconstrained_steepest_descent}
    \phi^{(n+1)}
    =
    \phi^{(n)}
    -
    \alpha_n \nabla \mathcal E \left( \phi^{(n)} \right)
    ,
\end{equation}
where $\alpha_n > 0$ are adaptively chosen steps \cite{Nocedal_Wright_book}.
Note that $- \nabla \mathcal E \left( \phi^{(n)} \right)$ is pointing
in the direction of the steepest descent of the energy $\mathcal E$
at the point $\phi^{(n)}$.
This particular form \eqref{unconstrained_steepest_descent} is valid
for the unconstrained optimization,
making it suitable for the orbital rotation formalism.
Roughly speaking, the gradient descent is comparable
in efficiency to the fixed-point iteration.
Furthermore, it is more robust, specifically when it comes to
the initial guess determination,
which is the cornerstone of the \ac{SCF} procedure \cite{Lehtola2019}.
Below I will demonstrate the convergence of the gradient descent
from a random initial guess.
Nevertheless, this method seems unfairly neglected in quantum chemistry,
even though it is the foundation of the quasi-Newton algorithms
used in the multiconfigurational problems \cite{Helgaker_Jorgensen_Olsen}.
As already mentioned above,
the standard \ac{SCF} is a well preconditioned fixed point scheme by default,
while Formula \eqref{unconstrained_steepest_descent} lacks a preconditioner.
In particular, in this work I show how to precondition the gradient descent
in a similar fashion to \eqref{preconditioned_nonlinear_eigenvalue_problem},
which makes it comparable in the amount of iterations needed to
the scheme \eqref{preconditioned_SCF_procedure}.

While the unconstrained gradient descent \eqref{unconstrained_steepest_descent},
is applicable in the orbital rotation representation
with $\kappa^{(n)}$ staying instead of $\phi^{(n)}$ in  
\eqref{unconstrained_steepest_descent}, of course.
It is not obvious how to extend this unconstrained picture to other
discretizations, grid-based and adaptive methods, in particular.
Scheme \eqref{unconstrained_steepest_descent} is impractical to use
in the Euler-Lagrange formalism with $\mathcal L$ staying instead of $\mathcal E$,
since all its stationary points \eqref{Euler_Lagrange_formulation}
are saddle points.
Moreover, working with rotations $\kappa$ instead of molecular orbitals $\phi$
can be tedious.

In their foundational paper, Edelman, Arias, and Smith \cite{Edelman_Arias_Smith}
identified electronic-structure theory as a Riemannian optimization problem
on the Stiefel manifold.
This perspective was subsequently developed primarily within the mathematical optimization community.
Absil, Mahony, and Sepulchre \cite{Absil2008} established the general framework
of optimization algorithms on matrix manifolds,
including precise definitions of gradients, retractions, vector transports, and convergence guarantees.
Their monograph has since become the standard reference in applied mathematics
and popularized Riemannian optimization as a method of choice
for constrained minimization in many fields.

Direct minimization algorithms for Kohn-Sham \ac{DFT} were formalized by
Schneider, Rohwedder, Neelov, and Blauert \cite{Schneider_Rohwedder_Neelov_Blauert2009},
who incorporated orthogonality constraints via manifold methods and provided both theoretical analysis and practical techniques for reliably computing occupied subspaces.
Zhang, Zhu, Wen, and Zhou \cite{Zhang_Zhu_Wen_Zhou2014} carried out a comprehensive analysis of gradient optimization schemes, including global and local convergence rates, and demonstrated clear computational advantages over SCF.
Building on this, Dai, Liu, Zhang, and Zhou \cite{Dai_Liu_Zhang_Zhou2017} leveraged the Stiefel manifold structure to design practical, parallelizable optimization algorithms, introducing efficient update strategies for large-scale atomic and molecular systems.
More recently, Dai, Gironcoli, Yang, and Zhou \cite{Dai_Gironcoli_Yang_Zhou2023}
advanced Riemannian conjugate-gradient and Newton-type algorithms for Kohn-Sham equations,
incorporating adaptive preconditioning and sophisticated metric choices to achieve both high performance and strong theoretical guarantees.

Recent work has extended these methods beyond matrix-based discretizations.
Luo, Wang, and Ren \cite{Luo_Wang_Ren2025} developed scalable solvers for Kohn-Sham equations in the finite-element setting,
using inexact Newton steps that respect the Stiefel structure.
Altmann, Peterseim, and Stykel \cite{Altmann_Peterseim_Stykel2024} introduced Riemannian optimization on generalized oblique manifolds,
with applications to multicomponent Bose-Einstein condensates,
providing energy-adaptive metrics and rigorous convergence proofs in Hilbert-space settings.
Building on this direction, Peterseim, Püschel, and Stykel \cite{Peterseim_Puschel_Stykel2025} specialized these ideas to Kohn-Sham \ac{DFT} for non-metallic crystals,
introducing an energy-adaptive metric that accelerates Riemannian conjugate gradient methods
and delivers competitive performance compared to SCF.

In quantum chemistry,
Vidal, Nottoli, Lipparini, and Cances \cite{Vidal_Nottoli_Lipparini_Cances2024}
analyzed orbital optimization for the \ac{ROHF} and \ac{CASSCF} methods
within a Riemannian framework,
showing that the relevant orbitals form a quotient (flag) manifold.
Their benchmarks demonstrate superior robustness of Riemannian algorithms in strongly correlated and symmetry-breaking cases,
highlighting automatic constraint enforcement and reduced parameter tuning.

The purpose of the current work is to provide a comprehensive description of
the Riemannian optimization for the Hartree-Fock and Kohn-Sham energy functionals
defined on the Stiefel manifold $\St(N)$
of the orthogonality constraint \eqref{orthonormality_constraint}.
This manifold is smoothly embedded into the Sobolev space $H^1$,
making it possible to define first the Euclidean gradient
$\nabla \mathcal E(\phi)$ for every orbital vector $\phi \in H^1$.
Then for $\phi \in \St(N)$ one can project it to the tangent space $T_{\phi}$
and obtain the corresponding Riemannian gradient $\grad \mathcal E(\phi)$.
In exactly the same manner as for the unconstrained problem,
$- \grad \mathcal E(\phi)$ corresponds to the direction
of the steepest decrease of $\mathcal E$ at $\phi$.
This will allow us to formalize the classical algorithm \eqref{unconstrained_steepest_descent}
for the general functional formulation of the Hartree-Fock minimization problem.
Then its preconditioned conjugate version is considered,
as a practical scheme.
Factoring out the orbital rotation gauge invariance of the energy functional
one introduces the Grassmann manifold $\Gr(N) = \St(N) / O(N)$,
the quotient of the surface $\St(N)$
with respect to the group $O(N)$ of orthogonal matrices of size $N \times N$.

All the numerical experiments are conducted using multiwavelet machinery
\cite{Alpert1993, Alpert_Beylkin_Gines_Vozovoi2002}, though
I do not see any theoretical restrictions on using other discretization techniques.
To the best of the author's knowledge, neither $H^1$ optimization viewpoint nor
the multiwavelet version of gradient descent have ever been implemented.
In order to make the exposition accessible to a wider range of readers,
we start the Riemannian optimization for two-electron systems,
which corresponds to $N = 1$.
In this simple case the Stiefel manifold $\St(1)$ coincides with the $L^2$-sphere
embedded in the $H^1$ space of orbitals.
The Hartree-Fock theory for many electron systems
on the general manifolds $\St(N)$ and $\Gr(N)$
follows in the subsequent sections.

Finally, we can conclude the introduction by pointing out the
two most crucial advantages of the Riemannian optimization
of the electronic structure, in author's opinion:
it is flexible and robust.
The flexibility is supported by the opportunity of using different
discretization techniques including multiwavelets.
The robustness is supported by the convergence from the random initial data,
which neither \ac{SCF} nor its \ac{DIIS} accelerated version demonstrate.
Moreover, the author believes that the Riemannian gradient descent
is relatively easy to implement and it has the potential to outperform
the state of the art of \ac{DIIS}.
The latter, however, cannot be supported at the current stage.

\section{Two-electron formulation}
\label{Two_electron_formulation_section}
\setcounter{equation}{0}

This section primarily serves an illustrative purpose.
It also introduces the crucial concepts of Riemannian optimization,
which are the infinite dimensional extensions
of the basic finite dimensional geometric tools \cite{Boumal2023}.
Considered herein is the Hartree-Fock energy functional
\begin{equation}
\label{helium_energy}
    \mathcal E(\phi)
    =
    \norm{\nabla \phi}_{L^2}^2
    +
    2
    \Innerprod{V \phi}{\phi}_{L^2}
    +
    \Innerprod{J \left( \phi^2 \right)}{\phi^2}_{L^2}
\end{equation}
for a two-electron system,
defined on the Sobolev space $H^1 \left( \mathbb R^3 \right)$.
By $J$ one denotes convolution with $1/|x|$,
an integral operator that typically appears in Coulomb-exchange terms of \ac{SCF} expressions.
Formally, the Coulomb convolution is $J = 4\pi R(0)$,
where $R(0)$ is the limiting case for the resolvent of Laplacian $- \Delta$
defined by
\begin{equation}
\label{resolvent}
    R(-\mu) \phi
    =
    (-\Delta + \mu)^{-1} \phi
    =
    \frac{e^{- \sqrt{\mu} |x|}}{4 \pi |x|} * \phi
    , \quad
    \mu > 0
    .
\end{equation}
One shortly writes $R = R(-1)$ for $\mu = 1$, which appears in the expressions below,
due to the standard choice of inner product \eqref{Sobolev_inner_product}.

In order to describe the geometry of
the orthogonality constraint \eqref{orthonormality_constraint},
it is convenient to introduce a functional $\mathcal Q$
defined on the entire space $H^1$ and such that \eqref{orthonormality_constraint}
reads as
\begin{equation}
\label{definition_Q_helium}
    \mathcal Q(\phi)
    =
    \norm{\phi}_{L^2}^2
    =
    1
\end{equation}
for $N = 1$.
Note that with respect to the metric embedded from $H^1 \left( \mathbb R^3 \right)$
it is not a sphere.
Nevertheless, this surface is smooth and the functional $\mathcal Q$ is differentiable.
Indeed, by the linearization argument, if the derivative $d\mathcal Q(\phi)$ exists
then it satisfies
\[
    d\mathcal Q(\phi)(\delta \phi)
    =
    2 \int \phi \delta \phi
    , \quad
    \text{ for any }
    \,
    \delta \phi \in H^1 \left( \mathbb R^3 \right)
    .
\]
This line, obviously, defines a bounded linear functional
over the space of orbital updates
\(
    \delta \phi \in H^1 \left( \mathbb R^3 \right)
    ,
\)
with the operator norm
\(
    \norm{
        d\mathcal Q(\phi)
    }
    \leqslant
    2
    \norm{
        \phi
    }_{L^2}
    \leqslant
    2
    \norm{
        \phi
    }_{H^1}
    .
\)
Every bounded linear functional over a Hilbert space
can be uniquely represented by a particular element in this space.
In the case of the derivative $d\mathcal Q(\phi)$
acting over the Hilbert space $H^1 \left( \mathbb R^3 \right)$
this element, denoted by $\nabla \mathcal Q(\phi)$,
is called the Euclidean gradient.
This representation means that for any
\(
    \delta \phi \in H^1 \left( \mathbb R^3 \right)
\)
the following holds true
\[
    \Innerprod{ \nabla \mathcal Q(\phi) }{ \delta \phi }_{H^1}
    =
    d\mathcal Q(\phi)(\delta \phi)
    =
    2 \Innerprod{\phi}{\delta \phi}_{L^2}
    .
\]
In general, if
\(
    \Innerprod{ g }{ \delta \phi }_{H^1}
    =
    \Innerprod{f}{\delta \phi}_{L^2}
\)
for any 
\(
    \delta \phi \in H^1 \left( \mathbb R^3 \right)
    ,
\)
then this equality holds for each twice differentiable function
\(
    \delta \phi \in H^2 \left( \mathbb R^3 \right)
    ,
\)
in particular.
Therefore,
integrating by parts one can rewrite this equality
in the form
\(
    \Innerprod{ g }{ ( - \Delta + 1 ) \delta \phi }_{L^2}
    =
    \Innerprod{f}{\delta \phi}_{L^2}
    .
\)
Now, by changing variable $\delta \phi = R \delta \psi$
with the resolvent $R = ( - \Delta + 1 )^{-1}$ and
$\delta \psi \in L^2 \left( \mathbb R^3 \right)$,
one arrives at
\(
    \Innerprod{ g }{ \delta \psi }_{L^2}
    =
    \Innerprod{Rf}{\delta \psi}_{L^2}
    .
\)
Accounting for the arbitrariness of $\delta \psi$
one deduces that $g = Rf$ and
\begin{equation}
\label{nabla_Q_helium}
    \nabla \mathcal Q(\phi)
    =
    2 R \phi
    =
    2
    ( - \Delta + 1 )^{-1} \phi
    \in H^3 \left( \mathbb R^3 \right) \subset H^1 \left( \mathbb R^3 \right)
    ,
\end{equation}
in particular.

The tangent space $T_{\phi}$ at $\phi$ is defined as the kernel of $d\mathcal Q(\phi)$,
namely,
\[
    T_{\phi}
    =
    \left \{
        \delta \phi \in H^1 \left( \mathbb R^3 \right)
        :
        \Innerprod{\delta \phi}{\phi}_{L^2} = 0
    \right\}
    .
\]
Let me point out
that, in general $\delta \phi \in T_{\phi}$
does not mean that $\delta \phi$ is orthogonal to $\phi$,
since orthogonality refers to $H^1$-orthogonality by default.
In other words,
the tangent space $T_{\phi}$
coincides with the subspace of functions that are $H^1$-orthogonal to $\nabla \mathcal Q(\phi)$,
and the corresponding projection has the form
\[
    \Proj_{\phi} u
    =
    u
    -
    \frac
    {
        \Innerprod{ u }{ \nabla \mathcal Q(\phi) }_{H^1}
    }
    {
        \norm{\nabla \mathcal Q(\phi)}_{H^1}^2
    }
    \nabla \mathcal Q(\phi)
    , \quad
    \text{ for }
    \,
    u \in H^1 \left( \mathbb R^3 \right)
    .
\]
Substituting \eqref{nabla_Q_helium} instead of $\nabla \mathcal Q(\phi)$
one can rewrite the projection as
\begin{equation}
\label{projection_helium}    
    \Proj_{\phi} u
    =
    u
    -
    \frac
    {
        \Innerprod{ u }{\phi}_{L^2}
    }
    {
        \Innerprod{R \phi}{\phi}_{L^2}
    }
    R \phi
    , \quad
    \text{ for }
    \,
    u \in H^1 \left( \mathbb R^3 \right)
    .
\end{equation}

The energy functional \eqref{helium_energy} has the following derivative
\begin{equation}
\label{helium_derivative_E}
    d \mathcal E(\phi)(\delta \phi)
    =
    2
    \Innerprod{\nabla \phi}{\nabla \delta \phi}_{L^2}
    +
    4
    \Innerprod{V \phi}{\delta \phi}_{L^2}
    +
    4
    \Innerprod{J \left( \phi^2 \right)}{\phi \delta \phi}_{L^2}
    .
\end{equation}
The right hand side can be rewritten in terms of
the Sobolev inner product \eqref{Sobolev_inner_product},
which gives
\[
    d \mathcal E(\phi)(\delta \phi)
    =
    \Innerprod
    {
        2 \phi
        +
        4
        ( - \Delta + 1 )^{-1}
        \left(
            \left(
                V + J \left( \phi^2 \right) - \frac 12
            \right)
            \phi
        \right)
    }
    {\delta \phi}_{H^1}
    ,
\]
and so one obtains the Euclidean gradient
\begin{equation}
\label{helium_nabla_E}
    \nabla \mathcal E(\phi)
    =
    2 \phi
    +
    4
    R
    \left(
        \left(
            V + J \left( \phi^2 \right) - \frac 12
        \right)
        \phi
    \right)
\end{equation}
by the arbitrariness of the function 
\(
    \delta \phi \in H^1 \left( \mathbb R^3 \right)
    .
\)
In this paper we work in the embedded geometry setting,
which means that one considers the tangent space $T_{\phi}$
as a subspace of the Hilbert space $H^1 \left( \mathbb R^3 \right)$
with the inner product \eqref{Sobolev_inner_product}.
In particular, it implies that a Riemannian gradient $\grad \mathcal E(\phi)$
coincides with the projection of the corresponding Euclidean gradient:
\(
    \grad \mathcal E(\phi)
    =
    \Proj_{\phi} \nabla \mathcal E(\phi)
    .
\)
Therefore, the Riemannian gradient has the form
\begin{equation}
\label{helium_grad_E}
    \grad \mathcal E(\phi)
    =
    \nabla \mathcal E(\phi)
    -
    \frac
    {
        \Innerprod{\nabla \mathcal E(\phi)}{\phi}_{L^2}
    }
    {
        \Innerprod{R \phi}{\phi}_{L^2}
    }
    R \phi
\end{equation}
with the Euclidean gradient $\nabla \mathcal E(\phi)$
given by \eqref{helium_nabla_E}.

We are heading to a generalization of
the descent algorithm \eqref{unconstrained_steepest_descent}.
Nevertheless, it is worth drawing a connection to the fixed-point iteration
alternative \eqref{preconditioned_SCF_procedure}.
The extrema satisfy
the stationarity equation
\(
    \grad \mathcal E(\phi) = 0
    ,
\)
which is equivalent to the Lagrangian formulation
\(
    \nabla \mathcal E(\phi)
    =
    \lambda(\phi) \nabla \mathcal Q(\phi)
    ,
\)
by \eqref{nabla_Q_helium}, \eqref{helium_grad_E}.
It can be rewritten in the following form
\[
    \phi
    =
    -
    2 R
    \left(
        \left(
            V + J \left( \phi^2 \right)
        \right)
        \phi
        -
        \frac {1 + \lambda}2 \phi
    \right)
    .
\]
Using the Hilbert identity
\[
    R = R(\lambda) - (1 + \lambda) R(\lambda) R
\]
one obtains
\[
    \phi
    =
    -
    2 R(\lambda)
    \left(
        \left(
            V + J \left( \phi^2 \right)
        \right)
        \phi
        -
        \frac {1 + \lambda}2 \phi
    \right)
    +
    2 (1 + \lambda) R(\lambda) R
    \left(
        \left(
            V + J \left( \phi^2 \right)
        \right)
        \phi
        -
        \frac {1 + \lambda}2 \phi
    \right)
    ,
\]
which simplifies further to
\[
    \phi
    =
    -
    2 R(\lambda)
    \left(
        \left(
            V + J \left( \phi^2 \right)
        \right)
        \phi
    \right)
\]
that is equivalent to
the \ac{SCF} formulation \eqref{preconditioned_nonlinear_eigenvalue_problem}
with $\lambda = 2 \varepsilon$.
Notably, while deriving this expression we did not have to
appeal to the notion of generalized functions,
which one has to deal with
when deriving this expression from \eqref{nonlinear_eigenvalue_problem}.

The final piece we need in order to generalize
the gradient descent scheme \eqref{unconstrained_steepest_descent},
is a retraction $\mathcal R_{\phi}$,
a smooth map guaranteeing that during the optimization procedure
the orbitals $\phi^{(n)}$ stay on the $L^2$-sphere $\St(1)$
defined by \eqref{definition_Q_helium}.
The simplest and most practical example is the following
\begin{equation}
\label{retraction_helium}
    \mathcal R_{\phi}(v)
    =
    \frac
    {
        \phi + v
    }
    {
        \norm{\phi + v}_{L^2}
    }
    , \quad
    v \in T_{\phi}
    ,
\end{equation}
which is a retraction of at least first order.
In fact, due to the difference between $H^1$- and $L^2$-metrics,
it turns out to be exactly of the first order.
In other words,
this retraction does not yield the closest point on the manifold in $H^1$-metric,
in contrast to $L^2$-metric \cite{Carlson_Keller}.
Note that the corresponding finite dimensional analog is second order.
In practical calculations, however, this does not have any significant effect.

Finally, the Riemannian generalization of
the unconstrained steepest gradient descent \eqref{unconstrained_steepest_descent}
takes the form
\begin{equation}
\label{steepest_descent}
    \phi^{(n+1)}
    =
    \mathcal R_{\phi^{(n)}}
    \left(
        - \alpha_n \grad \mathcal E \left( \phi^{(n)} \right)
    \right)
    ,
\end{equation}
where $\alpha_n > 0$ are adaptively chosen steps,
similarly to the unconstrained optimization.
Notably,
the expressions \eqref{helium_nabla_E}, \eqref{helium_grad_E},
\eqref{retraction_helium}, \eqref{steepest_descent}
require neither explicit evaluation of derivatives $\partial_{x_i} \phi$ 
nor explicit $H^1$ inner products.
Apart from multiplication by the potential $V$,
the most expensive operations are Coulomb-type convolutions, which are particularly well suited to multiwavelet discretizations \cite{Harrison_Fann_Yanai_Beylkin2003}.
This motivates a paradigm shift: by representing orbitals in an adaptive multiwavelet basis, we can exploit the natural convolution structure of Riemannian gradients and achieve high accuracy at competitive computational cost.
With additional effort, other discretization schemes (plane waves, Gaussian basis sets, wavelets, or finite elements) can also be incorporated.

\begin{figure}[ht!]
\centering
\includegraphics[width=0.49\textwidth]
{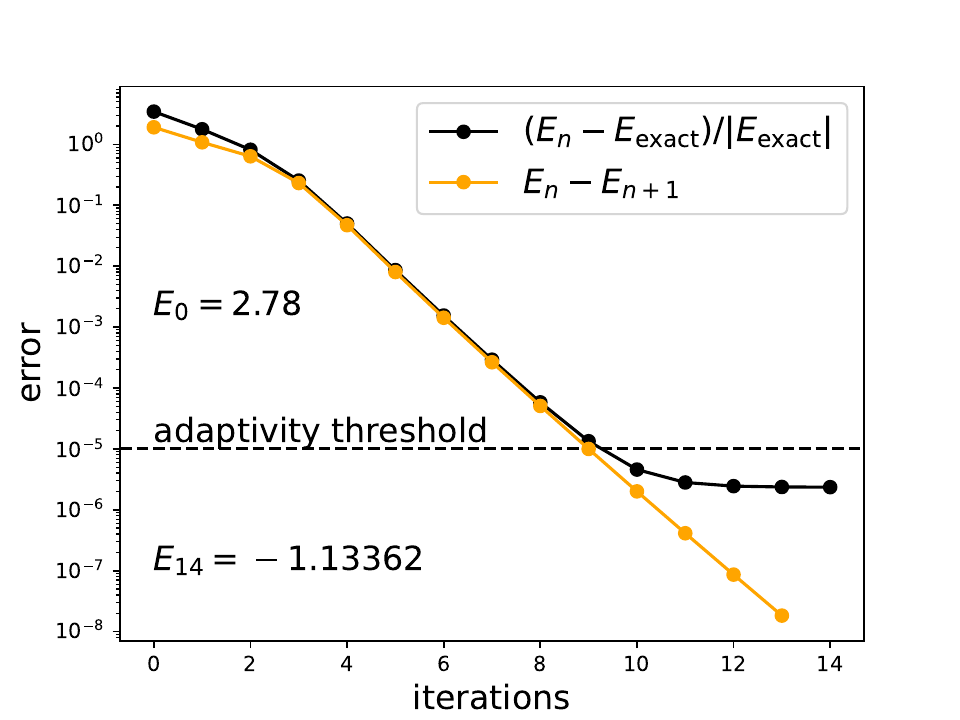}
\includegraphics[width=0.49\textwidth]
{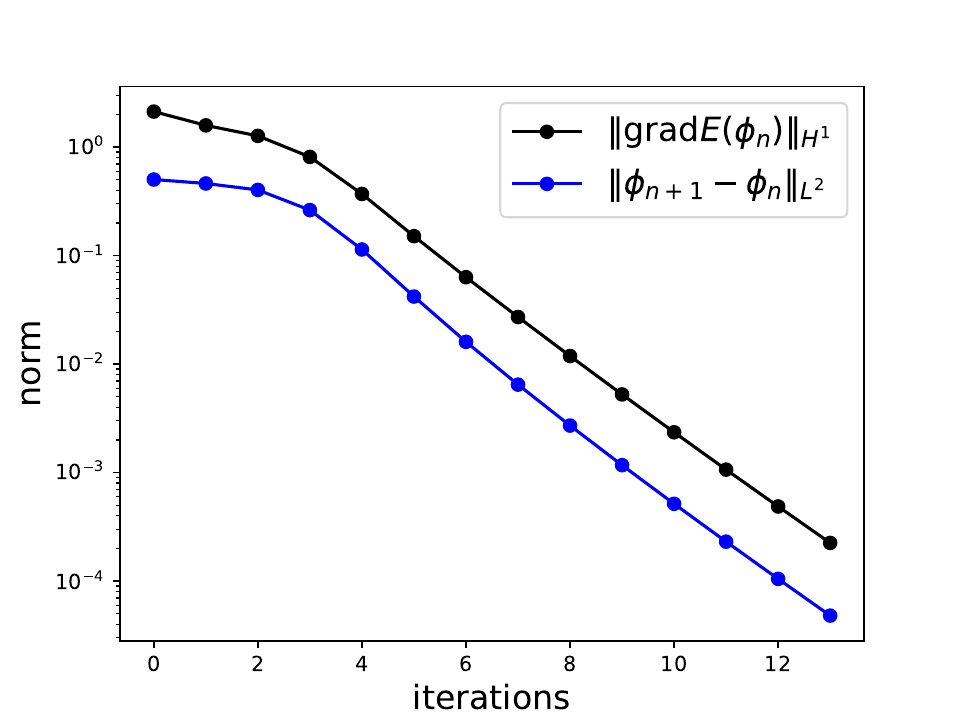}
\caption{
    Convergence of the Riemannian steepest gradient descent for H$_2$ Hartree-Fock model starting from a random Gaussian superposition.
    DIIS oscillates and converges in twice as many iterations.
}
\label{fig:h2_convergence}
\end{figure}

A common adaptive strategy to pick up the step-sizes $\alpha_n$
for the steepest gradient descent \eqref{steepest_descent}
is the so-called backtracking line-search.
It is convenient to introduce the function
\begin{equation}
\label{steepest_backtracking_g}
    g(\alpha)
    =
    \mathcal E
    \left(
        \mathcal R_{\phi}
        \left(
            - \alpha \grad \mathcal E(\phi)
        \right)
    \right)
\end{equation}
for a given fixed $\phi \in \St(1)$.
Clearly, $g(0) = \mathcal E(\phi)$.
Note that for $\phi = \phi^{(n)}$ and $\alpha = \alpha_n$
the value of this function $g(\alpha)$ coincides with
the energy $\mathcal E \left( \phi^{(n+1)} \right)$
at the next iterate $\phi^{(n+1)}$.
Therefore,
a satisfactory choice of step-size $\alpha_n$ should guarantee enough decrease of $g(\alpha)$.
During the backtracking the trial step-size is shrunk until
the acceptance condition of Armijo-Goldstein
\begin{equation}
\label{steepest_acceptance_condition}
    g(0) - g(\alpha)
    \geqslant
    r \alpha
    \norm{ \grad \mathcal E(\phi) }_{H^1}^2
\end{equation}
is satisfied for a given fixed parameter $r \in (0, 1)$,
in practice often set to $r = 10^{-4}$, see \cite{Nocedal_Wright_book}.
The shrinking is controlled by multiplying $\alpha$
with a constant $\tau \in (0, 1)$, normally set to $\tau = 1/2$.
Notice that overly aggressive shrinking may result in a lost opportunity.
Therefore, when the decrease of $g(\alpha)$ is large enough,
the first trial step-size at the next iteration is increased
by a growth factor $\gamma > 1$, say $\gamma = 1.2$, for instance.
The backtracking procedure takes place at each iteration
with given $\phi^{(n)}$, $\grad \mathcal E \left( \phi^{(n)} \right)$
and the first trial step $\bar \alpha_n$.
It is summarized in Algorithm \ref{algorithm_steepest_backtracking}.

\begin{algorithm}
\caption{Backtracking line-search for steepest descent}
\label{algorithm_steepest_backtracking}
    \begin{algorithmic}[1]
        \Parameters
        Constants $r, \tau \in (0,1)$, $\gamma > 1$,
        $r < \bar r < 1$
        and maximal step $\alpha_{\max} > 0$
        \Require
        Current iterate $\phi = \phi^{(n)} \in \St(1)$,
        corresponding $\grad \mathcal E \left( \phi \right)$,
        initial step $\bar\alpha_n > 0$
        \State Set $\alpha \gets \bar\alpha_n$
        \State
        Compute $g(\alpha)$
        by \eqref{steepest_backtracking_g}
        \While
        {
            \(
                g(0) - g(\alpha)
                <
                r \alpha
                \norm{ \grad \mathcal E(\phi) }_{H^1}^2
            \)
        }
            \State
            $\alpha \gets \tau \alpha$
            and recompute $g(\alpha)$
        \EndWhile
        \If
        {
            \(
                g(0) - g(\alpha)
                \geqslant
                \bar r \alpha
                \norm{ \grad \mathcal E(\phi) }_{H^1}^2
            \)
        }
            \State
            Choose new next trial step
            $\bar\alpha_{n+1} \gets \min(\gamma \alpha, \alpha_{\max})$
        \Else
            \State
            Keep accepted trial step $\bar\alpha_{n+1} \gets \alpha$
        \EndIf
        \State
        Set $\alpha_n \gets \alpha$
        \State
        Define next iterate $\phi^{(n+1)}$ by \eqref{steepest_descent}
    \end{algorithmic}
\end{algorithm}

Numerical experiments are performed using the multiwavelet machinery,
covered in
\cite{Alpert1993, Alpert_Beylkin_Gines_Vozovoi2002, Beylkin_Coifman_Rokhlin}
and often referred to as a \ac{MRA},
associated with wavelets \cite{Daubechies_book} in general.
Discretization of the convolution operators \eqref{resolvent}
is well described in \cite{Harrison_Fann_Yanai_Beylkin2003}.
Furthermore, an instructive pedagogical exposition can be found
in \cite{Frediani_Fossgaard_Fla_Ruud}.
An \ac{MRA} based computational software \mrchemsoft{}
is available at \cite{MRChemSoft}.
All numerical experiments presented here are conducted using this software.
Particularly, simulations of small molecules, including two-electron systems
under consideration,
are performed with the help of
the Python package \vampyr{} described in \cite{Bjorgve_2024_vampyr}.

The backtracking parameters are set to commonly used values:
\begin{equation}
\label{backtracking_parameters}
    r = 10^{-4}
    , \quad
    \bar r = 0.7
    , \quad
    \tau = \frac 12
    , \quad
    \gamma = 1.4
    , \quad
    \alpha_{\max} = 10
    .
\end{equation}
As an initial guess we take a collection of Gaussian functions
\begin{equation}
\label{random_gaussian_function}
    \phi_{\text{random}}(x)
    =
    \sum_{j=1}^{10}
    (-1)^j
    \exp\!\big(-|x - c_j|^2\big)
    ,
\end{equation}
with randomly distributed centers \(c_j \in \mathbb{R}^3\),
and normalize it as
\(
    \phi^{(0)}
    =
    \phi_{\text{random}} / \norm{ \phi_{\text{random}} }_{L^2}
    ,
\)
according to the requirement
\(
    \phi^{(0)}
    \in \St(1)
    .
\)
Setting the initial trial step $\bar \alpha_0 = 0.5$ we run
the Riemannian steepest gradient descent \eqref{steepest_descent}
following Algorithm \ref{algorithm_steepest_backtracking}.
For H$_2$ it converges to within \(10^{-5}\) relative accuracy,
set to be an \ac{MRA} threshold, 
in approximately 10 to 15 iterations,
see Figure~\ref{fig:h2_convergence}.
Remarkably, the simplest Riemannian optimizer without any preconditioning or acceleration outperforms DIIS,
that oscillates a lot and takes twice amount of iterations for the same random guess.
Notably, the gradient norm decreases monotonically to zero in the optimal way.  
For the energy reference $E_{\text{exact}}$
a high precision calculation was prepared with the default
\ac{SCF} scheme \eqref{preconditioned_SCF_procedure} of \mrchem{}.

As a matter of fact,
in the run presented in Figure \ref{fig:h2_convergence}
all the step sizes turned out to be the same $\alpha_n = \bar \alpha_0 = 0.5$
for all $n$.
For a well balanced set of parameters,
it is a common situation that the majority of trials are accepted.
In other words, the energy functional $\mathcal E$
is evaluated once per iteration normally.
In this particular run there were no rejections,
which supports a common practice of the gradient descent use.

Finally, the current section is concluded with the proof of
differentiability of \eqref{hartree_fock_energy}
and \eqref{helium_energy}, in particular.
This will justify the transition between
\eqref{helium_derivative_E} and \eqref{helium_grad_E}.
The claim is that Expression \eqref{helium_derivative_E}
defines a bounded linear functional with respect to
\(
    \delta \phi \in H^1 \left( \mathbb R^3 \right)
\)
for any given
\(
    \phi \in H^1 \left( \mathbb R^3 \right)
    .
\)
Clearly,
it is enough to demonstrate that there exists a positive constant $C$ such that
\begin{equation}
\label{nuclear_potential_energy_is_bounded}
    | (1 |V| 2) |
    =
    \left|
        \Innerprod{V \phi_1}{\phi_2}_{L^2}
    \right|
    \leqslant
    C
    \norm{ \phi_1 }_{H^1}
    \norm{ \phi_2 }_{H^1}
\end{equation}
and
\begin{equation}
\label{coulomb_exchange_energy_is_bounded}
    | (12|34) |
    =
    \left|
        \Innerprod{J \left( \phi_1 \phi_2 \right)}{\phi_3 \phi_4}_{L^2}
    \right|
    \leqslant
    C
    \norm{ \phi_1 }_{H^1}
    \norm{ \phi_2 }_{H^1}
    \norm{ \phi_3 }_{H^1}
    \norm{ \phi_4 }_{H^1}
\end{equation}
for arbitrary orbitals
\(
    \phi_j \in H^1 \left( \mathbb R^3 \right)
    .
\)

In order to prove \eqref{nuclear_potential_energy_is_bounded},
one may notice that a molecule can always be confined in
a bounded domain, say a ball $B_r = B_r(0) \subset \mathbb R^3$
with a large enough radius $r$ and a center at the origin.
This guarantees that all the Coulomb singularities are
strictly inside of this domain.
Then the potential $V$ is a square integrable function
inside of the ball $B_r$
and a bounded function outside $B_r' = \mathbb R^3 \setminus B_r$.
Hence splitting the $L^2$-inner product
in \eqref{nuclear_potential_energy_is_bounded}
into the integrals over $B_r$ and $B_r'$,
one deduces from the H\"older inequality that
\begin{multline*}
    | (1 |V| 2) |
    \leqslant
    \norm{ V }_{L^2(B_r)}
    \norm{ \phi_1 \phi_2 }_{L^2}
    +
    \norm{ V }_{L^{\infty}(B_r')}
    \norm{ \phi_1 \phi_2 }_{L^1}
    \\
    \leqslant
    \norm{ V }_{L^2(B_r)}
    \norm{ \phi_1 }_{L^4} \norm{ \phi_2 }_{L^4}
    +
    \norm{ V }_{L^{\infty}(B_r')}
    \norm{ \phi_1 }_{L^2} \norm{ \phi_2 }_{L^2}
    .
\end{multline*}
By the Sobolev embedding of $H^{s_p} \left( \mathbb R^3 \right)$
into $L^p \left( \mathbb R^3 \right)$
with $s_p \in [0, 3/2)$ such that $p = 6 / (3 - 2s_p)$,
we have
\[
    \norm{ \phi_j }_{L^4}
    \leqslant
    C_1
    \norm{ \phi_j }_{H^{3/4}}
    \leqslant
    C_1
    \norm{ \phi_j }_{H^1}
\]
and, of course,
\(
    \norm{ \phi_j }_{L^2}
    \leqslant
    \norm{ \phi_j }_{H^1}
\)
by the $H^1$-inner product definition \eqref{Sobolev_inner_product}.
Thus
\begin{equation*}
    | (1 |V| 2) |
    \leqslant
    \left(
        C_1^2
        \norm{ V }_{L^2(B_r)}
        +
        \norm{ V }_{L^{\infty}(B_r')}
    \right)
    \norm{ \phi_1 }_{H^1} \norm{ \phi_2 }_{H^1}
    .
\end{equation*}
The proof of \eqref{coulomb_exchange_energy_is_bounded}
follows from the H\"older inequality
and the Hardy-Littlewood-Sobolev fractional integration theorem
providing a bound for the convolution operator $J$
acting between $L^p \left( \mathbb R^3 \right)$
and  $L^q \left( \mathbb R^3 \right)$ with $p, q$ satisfying a specific relation.
Indeed, noticing
\[
    | (12|34) |
    \leqslant
    \norm{ J(\phi_1 \phi_2) }_{L^q}
    \norm{ \phi_3 \phi_4 }_{L^{q'}}
    \leqslant
    C_2
    \norm{ \phi_1 \phi_2 }_{L^p}
    \norm{ \phi_3 \phi_4 }_{L^{q'}}
    \leqslant
    C_2
    \norm{ \phi_1 }_{L^{2p}}
    \norm{ \phi_2 }_{L^{2p}}
    \norm{ \phi_3 }_{L^{2q'}}
    \norm{ \phi_4 }_{L^{2q'}}
\]
and appealing to the Sobolev embedding
with the new indices $s_{2p}, s_{2q'} \in [0, 3/2)$,
one can deduce
\[
    | (12|34) |
    \leqslant
    C_3
    \norm{ \phi_1 }_{H^{s_{2p}}}
    \norm{ \phi_2 }_{H^{s_{2p}}}
    \norm{ \phi_3 }_{H^{s_{2q'}}}
    \norm{ \phi_4 }_{H^{s_{2q'}}}
\]
as long as there exist positive real numbers satisfying
\[
    \frac 1p = \frac 1q + \frac 23
    , \quad
    \frac 1q + \frac 1{q'} = 1
    , \quad
    2p = \frac 6{ 3 - 2 s_{2p}}
    , \quad
    2q' = \frac 6{ 3 - 2 s_{2q'}}
    .
\]
Setting
\(
    p = q' = 6/5
    ,
    q = 6
    ,
    s_{2p} = s_{2q'} = 1/4
    ,
\)
for example,
one arrives at
\[
    | (12|34) |
    \leqslant
    C_3
    \norm{ \phi_1 }_{H^{1/4}}
    \norm{ \phi_2 }_{H^{1/4}}
    \norm{ \phi_3 }_{H^{1/4}}
    \norm{ \phi_4 }_{H^{1/4}}
    \leqslant
    C_3
    \norm{ \phi_1 }_{H^{1}}
    \norm{ \phi_2 }_{H^{1}}
    \norm{ \phi_3 }_{H^{1}}
    \norm{ \phi_4 }_{H^{1}}
    .
\]

It is a classical argument in the theory of
nonlinear partial differential equations.
The proofs of \eqref{nuclear_potential_energy_is_bounded}
and \eqref{coulomb_exchange_energy_is_bounded}
can likely be found elsewhere,
and therefore, they are presented here primarily for completeness.

\section{Conjugate gradient descent}
\label{Conjugate_gradient_descent_section}
\setcounter{equation}{0}

In practice the steepest descent algorithm is not used.
For larger molecules
one needs to use the conjugate gradient method with a restart procedure
\cite{Nocedal_Wright_book, Powell1977}.
It is also important to precondition the gradient,
as was mentioned in the introduction.
This section is devoted to the formulation of
the Riemannian nonlinear conjugate gradient descent,
applicable to
the energy optimization on
both Stiefel $\St(N)$ and Grassmann $\Gr(N)$ manifolds.
It is summarized in Algorithm \ref{algorithm_conjugate_gradient_descent},
divided into three logical parts:
line-search, generation of a conjugacy direction and a restart procedure.

\begin{algorithm}
\caption{Preconditioned conjugate gradient descent}
\label{algorithm_conjugate_gradient_descent}
    \begin{algorithmic}[1]
        \Parameters
        $\tau \in (0,1)$, $\gamma > 1$,
        $0 < r < \bar r < 1$, $\alpha_{\max} > 0$
        for backtracking
        and $\beta_{\max}, \eta^{\text{P}} > 0$, $n_{\text{res}} \in \mathbb N$
        for conjugacy and restart
        \Require
        Initial data $\phi^{(0)}$ with
        direction
        $p_0 = - \Prec_{ \phi^{(0)} } \grad \mathcal E \left( \phi^{(0)} \right)$
        and
        trial step $\bar\alpha_0 > 0$
        \For
        {
            $n = 0, 1, 2, \ldots$
            until convergence
        }
        \State
        Set $\alpha \gets \bar\alpha_n$
        and
        compute
        \(
            g(\alpha)
            =
            \mathcal E
            \left(
                \mathcal R_{ \phi^{(n)} }
                \left(
                    \alpha p_n
                \right)
            \right)
        \)
        \While
        {
            \(
                g(\alpha)
                >
                g(0)
                +
                r \alpha
                \Innerprod{p_n}{ \grad \mathcal E \left( \phi^{(n)} \right) }_{H^1}
            \)
        }
            \State
            $\alpha \gets \tau \alpha$
            and recompute $g(\alpha)$
        \EndWhile
        \If
        {
            \(
                g(\alpha)
                \leqslant
                g(0)
                +
                \bar r \alpha
                \Innerprod{p_n}{ \grad \mathcal E \left( \phi^{(n)} \right) }_{H^1}
            \)
        }
            \State
            Choose new next trial step
            $\bar\alpha_{n+1} \gets \min(\gamma \alpha, \alpha_{\max})$
        \Else
            \State
            Keep accepted trial step $\bar\alpha_{n+1} \gets \alpha$
        \EndIf
        \State
        Define next iterate
        $
            \phi^{(n+1)}
            =
            \mathcal R_{ \phi^{(n)} }
            \left(
                \alpha p_n
            \right)
        $
        \State
        Define
        \(
            \beta_n
            =
            \min
            \left \{
                \max
                \left \{
                    \beta_n^{\text{PR}}
                    ,
                    0
                \right\}
                ,
                \beta_{\max}
            \right\}
        \)
        with
        \begin{equation}
        \label{polak_ribiere}
            \beta_n^{\text{PR}}
            =
            \frac
            {
                \Innerprod
                {
                    \Prec_{ \phi^{(n+1)} }
                    \grad \mathcal E
                    \left( \phi^{(n+1)} \right)
                    -
                    \mathcal T_{ \phi^{(n+1)} \leftarrow \phi^{(n)} }
                    \Prec_{ \phi^{(n)} }
                    \grad \mathcal E
                    \left( \phi^{(n)} \right)
                }
                {
                    \grad \mathcal E
                    \left( \phi^{(n+1)} \right)
                }
                _{H^1}
            }
            {
                \Innerprod
                {
                    \Prec_{ \phi^{(n)} }
                    \grad \mathcal E
                    \left( \phi^{(n)} \right)
                }
                {
                    \grad \mathcal E
                    \left( \phi^{(n)} \right)
                }
                _{H^1}
            }
        \end{equation}
        \State
        Compute
        $
            p_{n+1}
            =
            -
            \Prec_{ \phi^{(n+1)} } \grad \mathcal E \left( \phi^{(n+1)} \right)
            +
            \beta_{n}
            \mathcal T_{ \phi^{(n+1)} \leftarrow \phi^{(n)} }
            \left(
                p_n
            \right)
        $
        \If
        {
            \(
                \Innerprod
                { p_{n+1} }
                { \grad \mathcal E \left( \phi^{(n+1)} \right) }_{H^1}
                \geqslant
                0
            \)
            or
            \(
                n - n( \text{previous restart} ) > n_{\text{res}}
            \)
            and
            \begin{equation}
            \label{powell_restart_condition}
                \frac
                {
                    \Innerprod
                    {
                        \mathcal T_{ \phi^{(n+1)} \leftarrow \phi^{(n)} }
                        \Prec_{ \phi^{(n)} }
                        \grad \mathcal E
                        \left( \phi^{(n)} \right)
                    }
                    {
                        \grad \mathcal E
                        \left( \phi^{(n+1)} \right)
                    }
                    _{H^1}
                }
                {
                    \Innerprod
                    {
                        \Prec_{ \phi^{(n)} }
                        \grad \mathcal E
                        \left( \phi^{(n)} \right)
                    }
                    {
                        \grad \mathcal E
                        \left( \phi^{(n)} \right)
                    }
                    _{H^1}
                }
                \geqslant
                \eta^{\text{P}}
            \end{equation}
        }
        \State
        Restart by setting
        \(
            n( \text{previous restart} ) \gets n
            ,
        \)
        $\beta_n \gets 0$
        and recomputing $p_{n+1}$ accordingly
        \EndIf
        \EndFor
    \end{algorithmic}
\end{algorithm}

Arguably, the most crucial part in determining the search direction
is preconditioning of the energy gradient.
A preconditioner $\Prec_{ \phi }$ is a positive definite linear operator
acting in the tangent space $T_{\phi}$.
A concrete working example is provided in the next section,
dealing with the electronic structure optimization for $2N$ electrons.
Here we just remark,
that the most general form of preconditioning in quantum chemistry
should factor out the kinetic energy at least,
as in Equation \eqref{preconditioned_nonlinear_eigenvalue_problem}.
Regardless of the particular discretization technique in use,
the Laplacian $\Delta$ is responsible for the stiffness of the problem,
or in other words, for the big condition number of the Hessian.
Physically speaking, the kinetic energy has the dominant effect
in the electronic structure.
As was pointed out in the introduction,
all \ac{SCF} schemes used in practice
include natural preconditioning.
As a matter of fact, it is not always straightforward
how to disable the default preconditioner.
While constructing a multiwavelet representation of the Laplacian $\Delta$
for dynamical problems in \cite{Dinvay2025heat},
I have demonstrated how catastrophic a discard
of the resolvent preconditioner in \eqref{preconditioned_SCF_procedure} can be:
increasing the number of iterations by four orders of magnitude.
Interestingly,
we do not have an explicit Laplacian term in \eqref{helium_nabla_E}.
Nevertheless, \eqref{helium_nabla_E} and \eqref{helium_grad_E}
are influenced by the kinetic energy,
and therefore, $\Prec_{ \phi }$ has to be introduced.

A line-search is conceptually similar to
Algorithm \ref{algorithm_steepest_backtracking},
with the main difference that the search direction
is no longer the steepest decrease of $\mathcal E\left( \phi^{(n)} \right)$,
but a conjugacy direction $p_n \in T_{ \phi^{(n)} }$.
It leads to a modification of
the acceptance condition of Armijo-Goldstein presented in
Algorithm \ref{algorithm_conjugate_gradient_descent},
compared to the corresponding
steepest version \eqref{steepest_acceptance_condition}.
The step-size $\alpha$ is adapted at each iteration
with the parameters defined in \eqref{backtracking_parameters}.
After each accepted step the step size is grown by the factor~$\gamma$
whenever the Armijo inequality is satisfied with sufficient margin.

At the first iteration one looks for the minimum
in the direction
$p_0 = - \Prec_{ \phi^{(0)} } \grad \mathcal E \left( \phi^{(0)} \right)$.
Later on, the conjugate vector $p_{n+1}$
is formed from a linear combination of the previous search direction $p_n$
and the preconditioned gradient
$\Prec_{ \phi^{(n+1)} } \grad \mathcal E \left( \phi^{(n+1)} \right)$.
Generally speaking, the tangent spaces $T_{ \phi^{(n+1)} }$ and $T_{ \phi^{(n)} }$
are different vector spaces.
Therefore,
a linear combination of their elements does not make sense.
In Riemannian optimization this technical difficulty is overcome by introducing
a notion of transporter $\mathcal T_{ \phi^{(n+1)} \leftarrow \phi^{(n)} }$,
a bounded linear operator acting between
$T_{ \phi^{(n)} }$ and $T_{ \phi^{(n+1)} }$.
Additionally \cite{Boumal2023},
it should depend smoothly on the orbitals $\phi^{(n)}, \phi^{(n+1)}$,
and furthermore,
it should coincide with the identity
whenever $\phi^{(n)} = \phi^{(n+1)}$.
It is not a uniquely defined notion,
and so
there are several possible options
\cite{Absil2008, Peterseim_Puschel_Stykel2025, Zhu2017, Zhu_Sato2020}
to choose from.
Moreover, there is a wide range of the conjugate parameter models for $\beta_n$,
see the survey \cite{Hager_Zhang2006survey}
for the unconstrained optimization.
With a bit of care these models extend to the Riemannian geometry.
Here one makes use of the Polak-Ribière model \eqref{polak_ribiere}
introduced in \cite{Polak_Ribiere1969, Polyak1969}.
The Polak-Ribière parameter is also capped,
$\beta_n \leqslant \beta_{\max}$ with $\beta_{\max} = 5$,
in order to prevent excessively large search directions.

Our conjugate gradient descent is also supplied with
the restart procedure,
which occasionally enforces $\beta_n = 0$
or equivalently,
\(
    p_{n+1}
    =
    -
    \Prec_{ \phi^{(n+1)} } \grad \mathcal E \left( \phi^{(n+1)} \right)
,
\)
in other words,
by taking the preconditioned steepest descent.
These restarts are performed independently on the computed values $\beta_n$.
This refreshes the algorithm, erasing the history
that may not be beneficial anymore.
Indeed, the conjugate gradient method is quite beneficial especially
in the quadratic region of $\mathcal E$ close to the minimum,
and in this domain it demonstrates superlinear convergence
provided one starts with the steepest step.
The information carried out from outside of this near minimum region,
may turn out to be harmful.
The conjugate gradient direction $p_{n+1}$ is restarted whenever
a descent-safeguard is violated.
Moreover, after $n_{\text{res}}$ iterations since the last restart
one checks if the two consecutive gradients are far from being orthogonal,
formalized by the Powell restart condition \eqref{powell_restart_condition}. 
We use a restart cooldown of four iterations, $n_{\text{res}} = 4$,
and a Powell threshold $\eta^{\text{P}}=0.3$.

Every tangent space $T_{\phi}$ is endowed with $H^1$-inner product.
It is natural, since we regard the Stiefel manifold $\St(N)$
as a surface in the Sobolev space in the embedded framework.
In the case of Grassmann manifold $\Gr(N)$ the meaning
of abused notation
\(
    \Innerprod{.}{.}_{H^1}
\)
will be clarified below in Section \ref{Grassmann_manifold_section}.
Algorithm \ref{algorithm_conjugate_gradient_descent}
is terminated
whenever the difference between the iterates
$\phi^{(n)}$ and $\phi^{(n+1)}$ becomes negligible.
Because of the considerable noise accompanying
multiwavelet calculations,
for termination the update is calculated using $L^2$-norm
instead of $H^1$.

\section{Stiefel manifold}
\setcounter{equation}{0}

In this section we formalize optimization on the Stiefel manifold $\St(N)$
in the infinite-dimensional setting of Sobolev orbitals.
We derive explicit expressions for the tangent space,
orthogonal projection, retraction,
and transporter compatible with the embedded $H^1$ geometry.
The resulting constructions generalize the two-electron case presented
above and form the foundation for the conjugate gradient algorithm.

Similarly to the two-electron problem considered in
Section \ref{Two_electron_formulation_section},
in order to describe the geometry of
the orthogonality constraint \eqref{orthonormality_constraint}
forming the orbital set $\St(N) \subset H^1$,
it is convenient to introduce a map
\(
    \mathcal Q : H^1 \to \mathbb R^{ N(N + 1) / 2 }
\)
having the following coordinate functionals
\begin{equation}
\label{definition_Q}
    \mathcal Q_{ij}(\phi)
    =
    \Innerprod{\phi_i}{\phi_j}_{L^2}
    , \quad
    \text{ for }
    \,
    i, j = 1, \ldots, N
    \,
    \text{ with }
    \,
    i \leqslant j
    .
\end{equation}
Thus \eqref{orthonormality_constraint}
reads as
\(
    \mathcal Q_{ij}(\phi)
    =
    \delta_{ij}
    .
\)
The map $\mathcal Q$ is differentiable and
\[
    \Innerprod{ \nabla \mathcal Q_{ij}(\phi) }{ \delta \phi }_{H^1}
    =
    d\mathcal Q_{ij}(\phi)(\delta \phi)
    =
    \Innerprod{\delta \phi_i}{\phi_j}_{L^2}
    +
    \Innerprod{\phi_i}{\delta \phi_j}_{L^2}
    , \quad
    \text{ for any }
    \,
    \delta \phi \in H^1
    \,
    \text{ and }
    \,
    i \leqslant j
    .
\]
The gradients
\(
    \nabla \mathcal Q_{ij}(\phi)
    =
    (
        \nabla_{\phi_1} Q_{ij}(\phi)
        , \ldots ,
        \nabla_{\phi_N} Q_{ij}(\phi)
    )^T
\)
can be found explicitly,
\[
    \nabla_{\phi_n} Q_{ij}(\phi)
    =
    \delta_{ni} R \phi_i
    + 
    \delta_{nj} R \phi_j
    .
\]
At $\phi \in \St(N)$
we define the tangent space
$T_{\phi} = \ker d \mathcal Q(\phi)$,
which coincides with the intersection of kernels
of $d\mathcal Q_{ij}(\phi)$.
Therefore,
$T_{\phi}$ can be characterized by orthogonality
to the gradients $\nabla \mathcal Q_{ij}(\phi)$.
In other words,
the projection $\Proj_{\phi}$ onto $T_{\phi}$
can be described
with the help of
\(
    R \phi
    =
    (
        R \phi_1
        , \ldots,
        R \phi_N
    )^T
    .
\)
Indeed, considering vectors $u, v \in H^1$ with $v = \Proj_{\phi} u$,
one may notice that $v = u - A R \phi$
with an $N \times N$ matrix $A$ depending on $u$.
Introducing two matrices $B_{ij} = \Innerprod{R \phi_i}{\phi_j}_{L^2}$
and $C_{ij} = \Innerprod{u_i}{\phi_j}_{L^2}$
we can reformulate the condition $v \in T_{\phi}$,
namely
\(
    \Innerprod{v_i}{\phi_j}_{L^2}
    +
    \Innerprod{\phi_i}{v_j}_{L^2}
    =
    0
    ,
\)
as
\[
    AB +(AB)^T = C + C^T
    .
\]
Hence the unknown matrix $A$ can be parametrised in the following way
\[
    A = (\Sym C + X)B^{-1}
    , \quad
    \Sym C = \frac{C + C^T}2
    , \quad
    X \text{ is skew symmetric.}
\]
It is clear that $B$ is symmetric.
Moreover, $B$ is strictly positive definite, and therefore
it is invertible.
Indeed,
for any vector $z \in \mathbb R^N$
we have the following quadratic form
\[
    z^T B z
    =
    \sum_{ij}
    z_i B_{ij} z_j
    =
    \Innerprod
    {
        \sum_i
        z_i
        \sqrt R \phi_i
    }
    {
        \sum_j
        z_j
        \sqrt R \phi_j
    }_{L^2}
    \geqslant
    0
    .
\]
Note that the equality is achieved only for $z = 0$,
since
\(
    \sum_i
    z_i
    \sqrt R \phi_i
    =
    0
\)
implies that 
\(
    \sum_i
    z_i
    \phi_i
    =
    0
\)
and the latter leads to $z = 0$
by the linear independence of orbitals $\phi_1, \ldots, \phi_N$.
The unknown matrix $X$ can be obtained from the minimization
of the norm
\(
    \norm{u - v}_{H^1}
\)
over
\(
    v \in T_{\phi}
    ,
\)
or equivalently,
by minimizing the following functional
\[
    \mathcal G(X)
    =
    \norm{AR \phi}_{H^1}^2
    =
    \Innerprod{AR \phi}{A \phi}_{L^2}
    =
    \trace A B A^T
    =
    \norm{A \sqrt B}_{\text{F}}^2
    =
    \norm{ (\Sym C + X) B^{-1/2} }_{\text{F}}^2
\]
over the linear space of skew symmetric matrices.
Here $\Innerprod{F}{G}_{\text{F}} = \trace F^T G$,
known as the Frobenius inner product of matrices $F$ and $G$.
By $\norm{F}_{\text{F}}$ we denote the norm induced by this product.
At the minimum we have
\(
    d \mathcal G(X) = 0
\)
implying
\[
    d \mathcal G(X)(\delta X)
    =
    2 \Innerprod{ \delta X }{ (\Sym C + X) B^{-1} }_{\text{F}}
    =
    0
\]
for all skew symmetric matrices $\delta X$.
Thus
\(
    A = (\Sym C + X)B^{-1}
\)
is symmetric.
This leads to
\begin{equation}
\label{sylvester_equation}
    AB + BA = C + C^T
    .
\end{equation}
In other words,
the projection is reduced to solving a Sylvester equation for
a symmetric $N \times N$ matrix $A$.
Equation \eqref{sylvester_equation}
uniquely determines $A$.
Many numerical linear algebra libraries provide routines for solving the Sylvester equation.
In this particular symmetric case,
the matrix $A$ can be found as
\(
    A = U Y U^T
    ,
\)
where the orthogonal transformation $U$ is diagonalizing $B = U \Lambda_B U^T$,
so that $\Lambda_B = \diag (\lambda_1, \ldots, \lambda_N)$,
and
\[
    Y_{ij} = \frac{ S_{ij} }{ \lambda_i + \lambda_j }
    , \quad
    S = U^T \left( C + C^T \right) U
    .
\]
We conclude the description of the tangent spaces
with the important formula
\begin{equation}
\label{projection}    
    \Proj_{\phi} u
    =
    u
    -
    A(u, \phi) R \phi
    , \quad
    \text{ for }
    \,
    u \in H^1
    ,
\end{equation}
generalizing \eqref{projection_helium} for $N \geqslant 1$.

For $2N$ electrons of a closed shell molecule
the energy functional $\mathcal E(\phi)$ has the form \eqref{hartree_fock_energy},
where
\begin{equation*}
    (i |h| j)
    =
    \frac 12
    \Innerprod{\nabla \phi_i}{\nabla \phi_j}_{L^2}
    +
    \Innerprod{V \phi_i}{\phi_j}_{L^2}
\end{equation*}
is a one-body integral
and
\begin{equation*}
    (ij|kl)
    =
    \Innerprod{J \left( \phi_i \phi_j \right)}{\phi_k \phi_l}_{L^2}
    =
    \int
    \frac{\phi_i(x) \phi_j(x) \phi_k(y) \phi_l(y)}{|x - y|}
    dxdy
\end{equation*}
is a two-body integral.
The derivative $d \mathcal E(\phi)$ of $\mathcal E(\phi)$
takes the value
\begin{equation}
\label{derivative_E}
    d \mathcal E(\phi)(\delta \phi)
    =
    4
    \sum_{i = 1}^N
    (i |h| \delta i)
    +
    4
    \sum_{i, j = 1}^N
    (
        2
        (ii|j \delta j)
        -
        (ij|i \delta j)
    )
\end{equation}
at an arbitrary orbital vector $\delta \phi$.
As above, from this expression one obtains
the Euclidean gradient $\nabla \mathcal E(\phi)$
consisting of the following components
\begin{equation*}
    \nabla_{\phi_i} \mathcal E(\phi)
    =
    2 \phi_i
    +
    4 R
    \left[
        \left(
            V - \frac 12
        \right)
        \phi_i
        +
        \sum_{j = 1}^N
        \left(
            2
            J(jj)
            \phi_i
            -
            J(ij)
            \phi_j
        \right)
    \right]
    .
\end{equation*}
Projecting it to the tangent space $T_{\phi}$ at $\phi \in \St(N)$,
$\grad \mathcal E(\phi) = \Proj_{\phi} \nabla \mathcal E(\phi)$,
one obtains the Riemannian gradient of
the Hartree-Fock energy functional \eqref{hartree_fock_energy}.
Thus
\begin{equation}
\label{grad_E}
    \grad \mathcal E(\phi)
    =
    T \phi
    +
    4 R V(\phi)
    ,
\end{equation}
where one has defined $T = 2 - (2 + A)R$
with the projection matrix $A = A(\nabla \mathcal E(\phi), \phi)$.
We refer to $T$ as the kinetic energy operator.
Following the discussion of the previous section
on importance of preconditioning out the kinetic part,
one may notice that
$T$ can be easily diagonalized together with $A = U \Lambda_A U^T$
and, as long as the spectrum of $A$ is negative, inverted as
\[
    T^{-1}
    =
    \frac 12
    U
    \left(
        1 + \left( 1 + \frac {\Lambda_A}2\right)
        R \left( \frac {\Lambda_A}2\right)
    \right)
    U^T
\]
allowing us to introduce a preconditioner as the following composition
\begin{equation*}
    \Prec_{\phi}
    =
    \Proj_{\phi} T^{-1}
    .
\end{equation*}
Importantly,
\(
    -
    \Prec_{\phi}
    \grad \mathcal E(\phi)
\)
is a descent direction,
due to the positivity of the quantity
\[
    \Innerprod
    {
        \Prec_{ \phi }
        \grad \mathcal E
        \left( \phi \right)
    }
    {
        \grad \mathcal E
        \left( \phi \right)
    }
    _{H^1}
    =
    \Innerprod
    {
        T^{-1}
        \grad \mathcal E
        \left( \phi \right)
    }
    {
        \grad \mathcal E
        \left( \phi \right)
    }
    _{H^1}
    =
    \norm
    {
        T^{-1/2}
        \grad \mathcal E
        \left( \phi \right)
    }
    _{H^1}^2
\]
away from a stationary point $\grad \mathcal E(\phi) = 0$.
In fact,
the right hand side can be estimated in terms of the gradient norm
a bit more precisely.
The Fourier symbol of a diagonal component of $T^{-1}$
associated to a negative eigenvalue $\mu \in \Spec A$
equals a function
\(
    \left(
        1
        +
        (1 + \mu / 2)
        \left(
            |\xi|^2 - \mu / 2
        \right)
    \right)
    / 2
\)
taking all possible values between
$1/2$ and $- 1 / \mu$
for
\(
    \xi \in \mathbb R^3
    .
\)
Therefore,
\[
    \min_{ \mu \in \Spec A }
    \left \{
        \frac 12
        ,
        - \frac 1{\mu}
    \right \}
    \leqslant
    \frac
    {
        \Innerprod
        {
            \Prec_{ \phi }
            \grad \mathcal E
            \left( \phi \right)
        }
        {
            \grad \mathcal E
            \left( \phi \right)
        }
        _{H^1}
    }
    {
        \norm
        {
            \grad \mathcal E
            \left( \phi \right)
        }
        _{H^1}^2
    }
    \leqslant
    \max_{ \mu \in \Spec A }
    \left \{
        \frac 12
        ,
        - \frac 1{\mu}
    \right \}
    .
\]

For a retraction $\mathcal R$
we use the following L{\"o}wdin orthonormalization procedure
\begin{equation}
\label{lowdin_retraction}
    \mathcal R_{\phi}(\delta \phi)
    =
    S^{-1/2}(\phi + \delta \phi)
    , \quad
    \delta \phi \in T_{\phi}
    ,
\end{equation}
where $S(\psi) = C(\psi, \psi)$ is the overlap matrix of the orbital vector
$\psi = \phi + \delta \phi$.
Recall that L{\"o}wdin transformation provides the closest
point on the Stiefel manifold $\St(N)$
in $L^2$-metric \cite{Carlson_Keller}, but not in $H^1$-metric.
Nevertheless,
it is a well defined first order retraction \cite{Altmann_Peterseim_Stykel2022},
serving all our needs.
For a transporter $\mathcal T_{ \phi \leftarrow \psi }$
with $\phi, \psi \in \St(N)$ we
make use of the projection \eqref{projection}
restricted to the tangent space $T_{\psi}$.
One can simply write
\(
    \mathcal T_{ \phi \leftarrow \psi }
    =
    \Proj_{\phi}
    .
\)
For a justification of this possible choice we
refer to \cite{Boumal2023},
where it is also explained the difference between
transporters and vector transports introduced in \cite{Absil2008}.
This completes the description of the Riemannian nonlinear
conjugate gradient descent algorithm \ref{algorithm_conjugate_gradient_descent}
for the Stiefel manifold $\St(N)$ of orthonormal orbitals.

\begin{figure}[ht!]
\centering
\includegraphics[width=0.49\textwidth]
{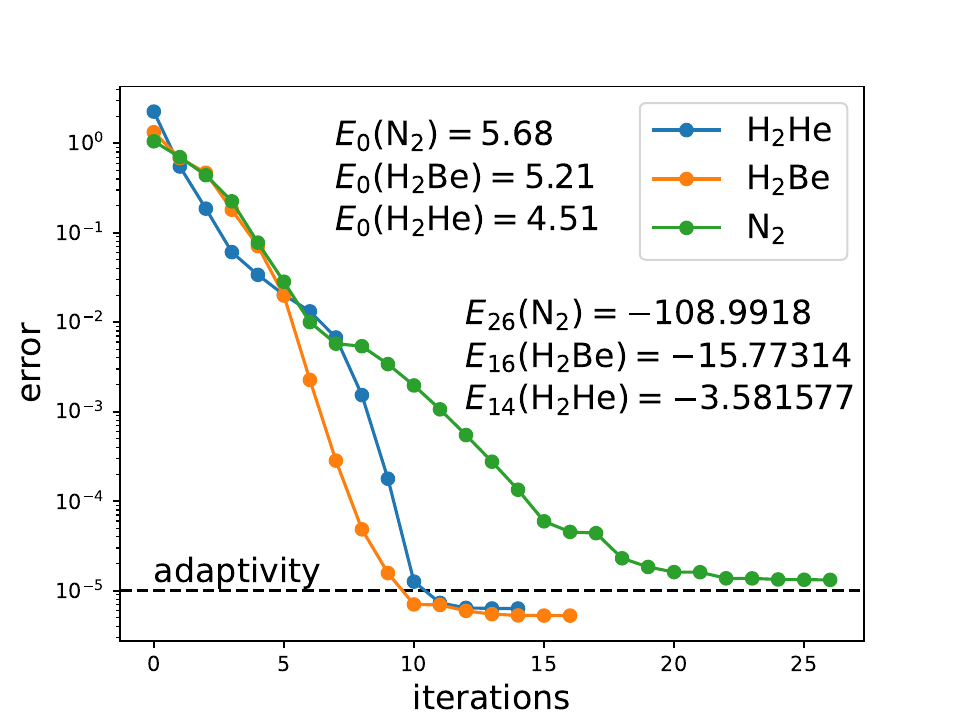}
\includegraphics[width=0.49\textwidth]
{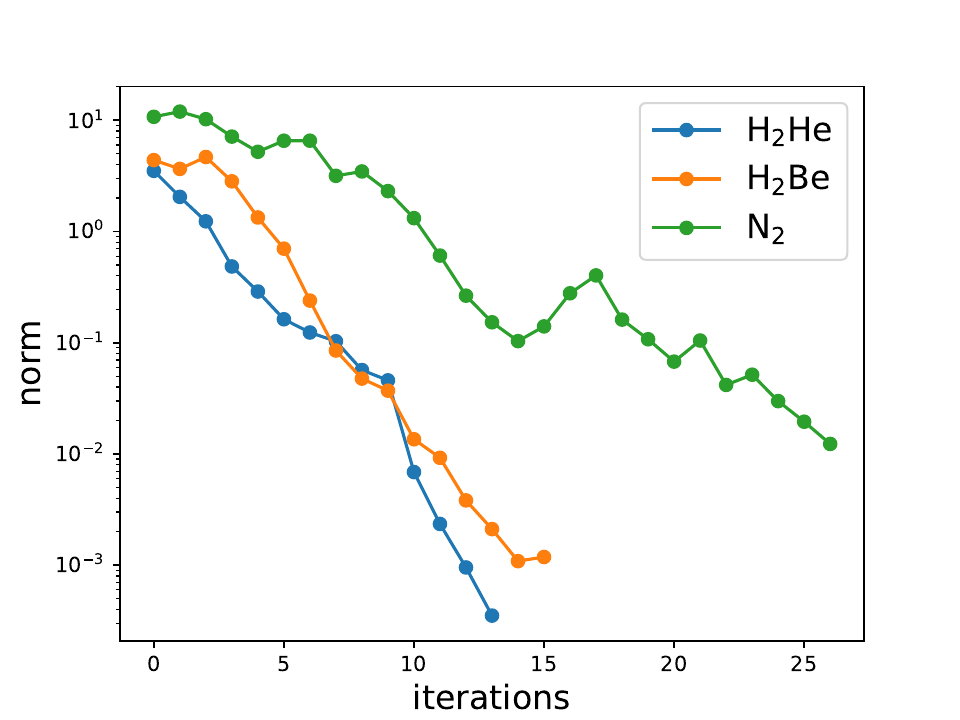}
\caption{
    Convergence of the Riemannian conjugate gradient descent
    on the Stiefel manifold for
    Hartree-Fock model starting from a random Gaussian superposition.
}
\label{fig:stiefel_random_convergence}
\end{figure}

Before moving further to numerical experiments
it is worth comparing with the fixed-point iteration
alternative \eqref{preconditioned_SCF_procedure}.
At the minimum we have $\grad \mathcal E(\phi) = 0$,
that is
\(
    \nabla \mathcal E(\phi)
    =
    A R \phi
    .
\)
Substituting the expression for $\nabla \mathcal E(\phi)$ one obtains
\[
    2 \phi - 2 R \phi + 4 R V(\phi)
    =
    A R \phi
    ,
\]
that is a Hartree-Fock system in $H^1$ space.
To see this one applies $R^{-1} = 1 - \Delta$ to both sides and arrives to
\[
    - \frac{\Delta}2 \phi + V(\phi)
    =
    \frac A4 \phi
    ,
\]
which coincides with the Hartree-Fock system written
in the standard form \eqref{nonlinear_eigenvalue_problem}
and interpreted as an equation in $H^{-1}$.
Thus at the minimum $A = 4 \varepsilon$,
that is, four times the Fock matrix.
The latter has negative eigenvalues,
associated with the occupied orbitals, of course.
In particular, the projection matrix $A(\nabla \mathcal E(\phi), \phi)$
has negative spectrum for $\phi$ close to the solution.
This justifies the kinetic energy operator inversion $T^{-1}$
and the definition of the preconditioner $\Prec_{\phi}$ introduced above.
Moreover, the numerical simulations conducted below
exhibit a negative spectrum of $A$ already after one or two iterations,
while starting from a random guess $\phi^{(0)}$.
For all chemically motivated initial data,
$A$ has only negative eigenvalues.

\begin{figure}[ht!]
\centering
\includegraphics[width=0.49\textwidth]
{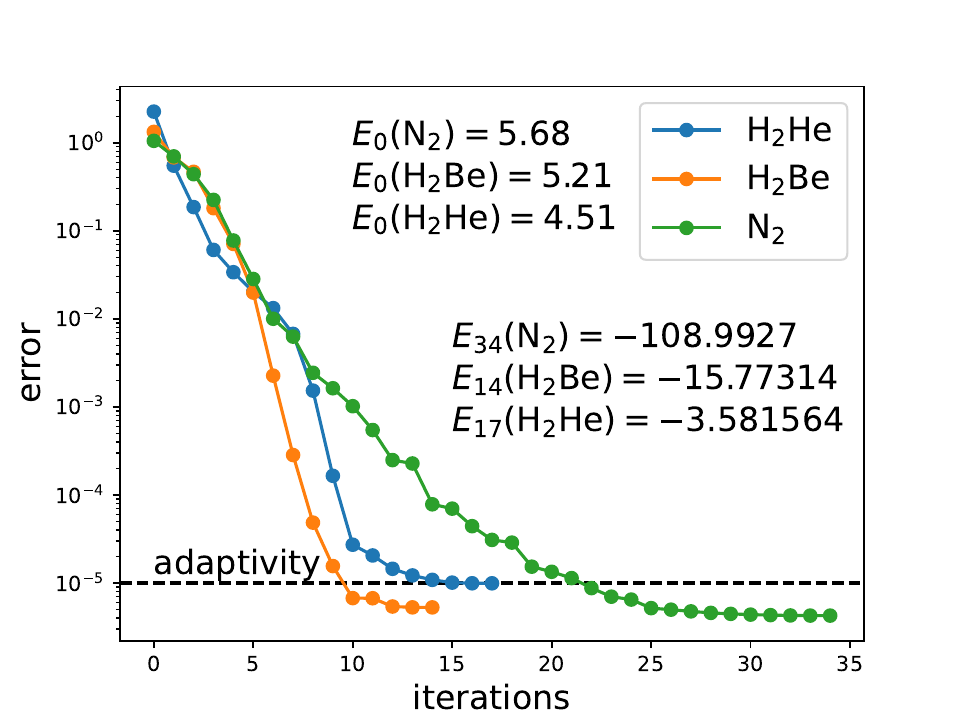}
\includegraphics[width=0.49\textwidth]
{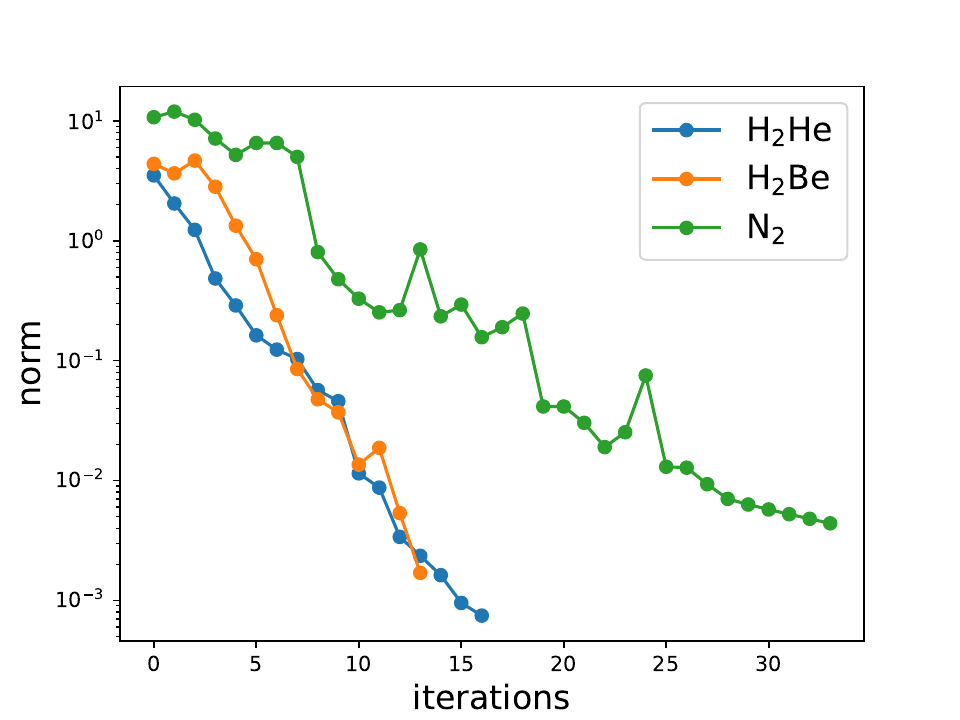}
\caption{
    Convergence of the Riemannian conjugate gradient descent
    on the Stiefel manifold for
    Hartree-Fock model starting from a random Gaussian superposition,
    with an additional restart after each rejection of the conjugate direction.
}
\label{fig:stiefel_random_convergence_modified}
\end{figure}

We first test our algorithm on small molecules,
starting from random initial data.
For $N = 2$ we take H$_2$He, which is a planar molecule
with each H--He bond equal to 1.81 bohrs
and a bond angle of $104.5^{\circ}$.
For $N = 3$ we consider H$_2$Be, which is a linear molecule
with a distance of 5.013 bohrs between the hydrogen atoms
and beryllium positioned in the middle.
For $N = 7$ we take N$_2$ with an N--N bond length of 2.074 bohrs.
Every coordinate $\phi_k^{(0)}$ of the initial guess $\phi^{(0)}$
is first initialized using \eqref{random_gaussian_function}.
Then $\phi^{(0)}$ orthonormalized by the L\"owdin transformation,
so that $\phi^{(0)} \in \St(N)$.
The backtracking parameters in use are \eqref{backtracking_parameters}
and the first trial step is set to $\bar \alpha_0 = 1.0$.
We run Algorithm \ref{algorithm_conjugate_gradient_descent} as described
in Section \ref{Conjugate_gradient_descent_section} 
and present the results
in Figure \ref{fig:stiefel_random_convergence}.

\begin{figure}[ht!]
\centering
\includegraphics[width=0.49\textwidth]
{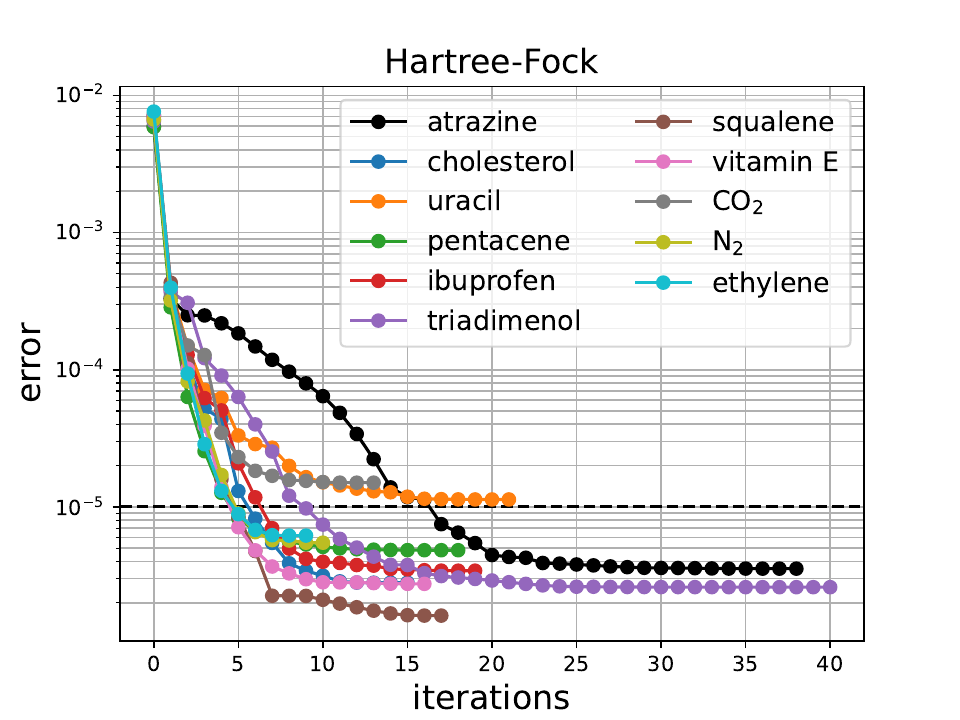}
\includegraphics[width=0.49\textwidth]
{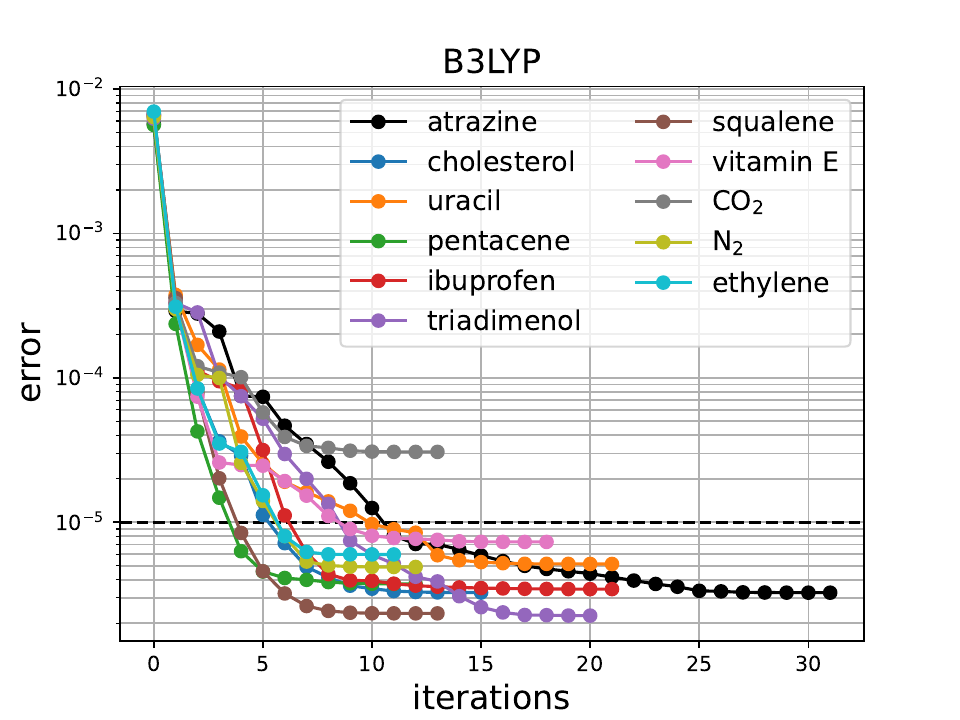}
\caption{
    Convergence of the Riemannian conjugate gradient descent
    on the Stiefel manifold for
    Hartree-Fock and B3LYP energy models.
}
\label{fig:stiefel_mrchem_energy_convergence}
\end{figure}

The energy error is calculated via
\(
    \left(
        \mathcal E
        \left(
            \phi^{(n)}
        \right)
        -
        \mathcal E_{\text{exact}}
    \right)
    / |\mathcal E_{\text{exact}}|
    ,
\)
where the reference value $\mathcal E_{\text{exact}}$
is obtained from high-precision calculations
in \mrchem{} with the state of the art \ac{SCF} algorithm.
In these particular calculations
\(
    \mathcal E
    \left(
        \phi^{(n)}
    \right)
    >
    \mathcal E_{\text{exact}}
    ,
\)
although this is not necessarily the case for other molecules.
The calculations in Figure \ref{fig:stiefel_random_convergence} are adaptive with 
the threshold set to $10^{-5}$.
This final numerical precision is the main reason for the slow down
of the algorithm when the error approaches this numerical barrier.
In normal \ac{SCF} calculations of \mrchem{} this problem is overcome
by making use of a dedicated \ac{DIIS} scheme named
\ac{KAIN} method \cite{Harrison_KAIN}
with a considerable history tracking, by default set to previous 5 iterations.
Algorithm \ref{algorithm_conjugate_gradient_descent} exploits only
the latest iterate for constructing conjugate direction.
Nevertheless,
it still demonstrates a rapid convergence away from the adaptivity threshold.
Moreover, it converges from random initial data,
whereas already for the molecules under consideration \ac{DIIS} fails
to converge.
The Riemannian energy gradient
$
    \grad \mathcal E
    \left(
        \phi^{(n)}
    \right)
$
tends to zero as one can see in Figure \ref{fig:stiefel_random_convergence},
where
$
    \norm{
        \grad \mathcal E
        \left(
            \phi^{(n)}
        \right)
    }_{H^1}
$
is depicted in the right panel.
Clearly, the corresponding curves cannot reach zero due to
a considerable amount of numerical noise accompanying \ac{MRA} in practice,
making the implementation of gradient descent challenging
in multiwavelets.

While the gradient descent performs well regardless of the initial data,
it is more affected by the noise caused by the adaptive calculations.
The method  described here is quite generic
concerning the discretization technique
and not tuned to
the multiwavelet framework, which will be done elsewhere.
Nevertheless,
the fact that in Figure \ref{fig:stiefel_random_convergence}
the green curve, associated with the nitrogen molecule,
does not cross the adaptivity threshold
is unsatisfactory.
This will be partially fixed in the next section,
where the redundant degrees of freedom are eliminated.
Alternatively,
we can also try to modify the restarting procedure of
Algorithm \ref{algorithm_conjugate_gradient_descent}.
As an example,
we restart after the first rejection during the backtracking.
This modification was used only in the calculations
presented in Figure \ref{fig:stiefel_random_convergence_modified}.
Its main purpose is to show,
that the generic Riemannian optimization can still be tuned to
noisy numerical techniques, specifically multiwavelet based.

\begin{figure}[ht!]
\centering
\includegraphics[width=0.49\textwidth]
{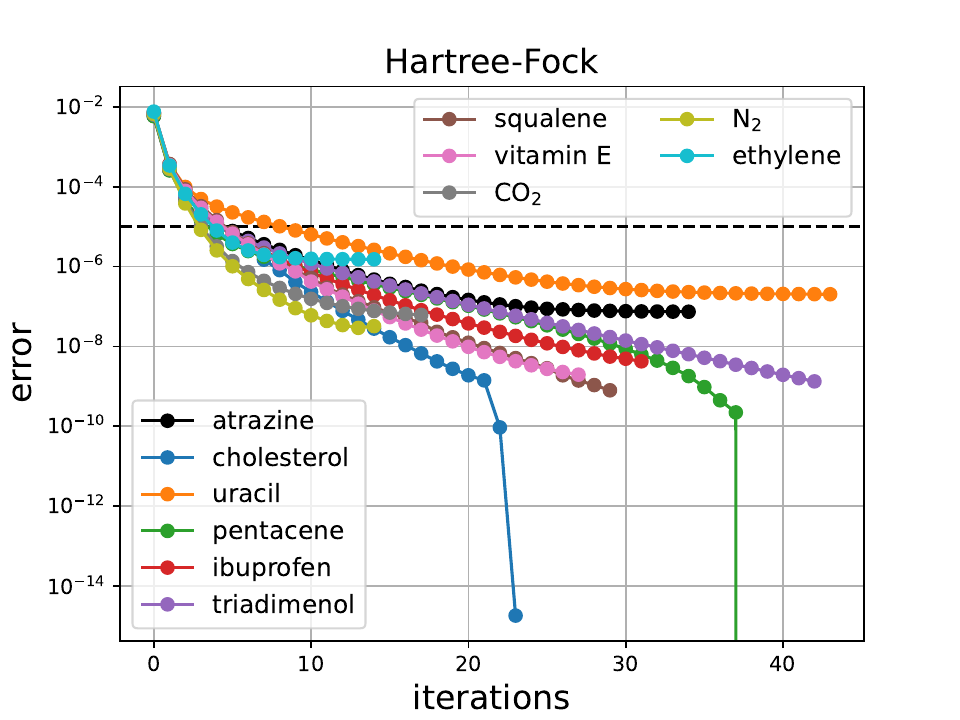}
\includegraphics[width=0.49\textwidth]
{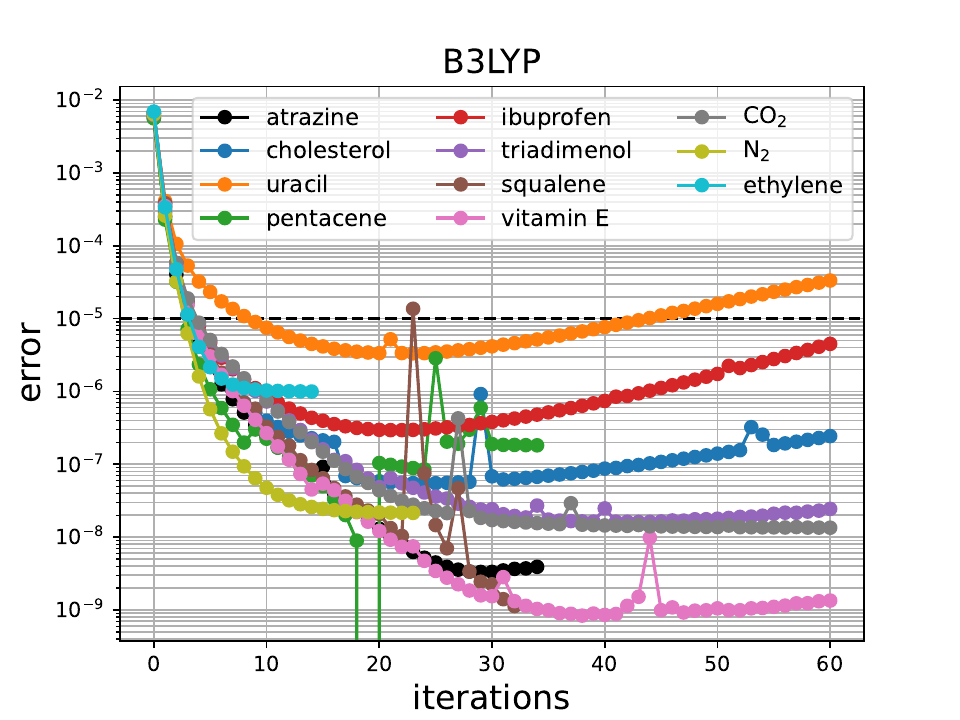}
\caption{
    Convergence of the \ac{SCF} method
    accelerated by \ac{KAIN} with one element in the history
    for Hartree-Fock and B3LYP energy models.
}
\label{fig:kain_1_mrchem_energy_convergence}
\end{figure}

We now turn to testing our \mrchem{} implementation of
Algorithm \ref{algorithm_conjugate_gradient_descent},
where the default initial guess construction is exploited.
The initial molecular orbitals are generated using the \ac{SAD} approach.
In this procedure the electron density of the molecule is approximated
by a sum of self-consistent atomic densities centered at the nuclear positions.
An effective Hartree-Fock operator constructed from this density
is diagonalized in a small Gaussian atomic-orbital basis (3-21G),
yielding approximate molecular orbitals.
These orbitals are subsequently projected onto the multiwavelet representation
used in the present calculations.

The orbital optimization is performed for
nitrogen
(N$_2$)
with $N = 7$,
ethylene
(C$_2$H$_4$)
with $N = 8$,
carbon dioxide (CO$_2$)
with $N = 11$,
uracil
(C$_4$H$_4$N$_2$O$_2$)
with $N = 29$,
ibuprofen
(C$_{13}$H$_{18}$O$_2$)
with $N = 56$,
atrazine
(C$_8$H$_{14}$ClN$_5$)
with $N = 57$,
pentacene
(C$_{22}$H$_{14}$)
with $N = 73$,
triadimenol
(C$_{14}$H$_{18}$ClN$_3$O$_2$)
with $N = 78$,
cholesterol
(C$_{27}$H$_{46}$O)
with $N = 108$,
squalene
(C$_{30}$H$_{50}$)
with $N = 115$
and
vitamin E
(C$_{29}$H$_{50}$O$_2$)
with $N = 120$.
The molecular geometries are taken from PubChem \cite{pubchem}.
For each molecule the reference energy 
\(
    \mathcal E_{\text{ref}}
\)
is defined as the lowest value obtained among several high-precision
\ac{SCF} calculations and the final gradient-descent iterate
\[
    \text{error}_n
    =
    \frac
    {
        \mathcal E
        \left(
            \phi^{(n)}
        \right)
        -
        \mathcal E_{\text{ref}}
    }
    {
        |\mathcal E_{\text{ref}}|
    }
    , \quad
    \mathcal E_{\text{ref}}
    =
    \min
    \left \{
        \mathcal E_{\text{SCF}}^{(k)}
        ,
        \mathcal E_{\text{GD}}^{(\text{final})}
    \right \}
    .
\]
For each molecule we optimize the Hartree-Fock energy \eqref{hartree_fock_energy}
and the B3LYP functional.
Note that for determining $\nabla \mathcal E(\phi)$ and
the Riemannian gradient \eqref{grad_E} one only needs $V(\phi)$,
which is obviously
defined for all the \ac{DFT} functionals used in the Kohn-Sham theory.
The results are presented in Figure \ref{fig:stiefel_mrchem_energy_convergence}.
For comparison we also run \ac{SCF} for the same molecules with \ac{KAIN}.
To make the comparison fair, we keep exactly one element in the \ac{KAIN} history,
which is consistent with transportation of one element while constructing
the conjugate direction in
the Riemannian algorithm \ref{algorithm_conjugate_gradient_descent}.

As was mentioned above, the gradient decent is more noise affected
compared with \ac{SCF}.
In Figure \ref{fig:stiefel_mrchem_energy_convergence}
the energy error curves for the carbon dioxide stop above the adaptivity threshold.
Also the Hartree-Fock curve for uracil stops just above the threshold.
Moreover, there are several long iteration tails below the threshold.
Note that \ac{SCF} has the long tails as well,
but they are all below the tolerance line
for the Hartree-Fock in Figure \ref{fig:kain_1_mrchem_energy_convergence}.
The convergence is assumed whenever the orbital update is less than $10^{-4}$,
that is ten times adaptivity parameter.
This criterion turns out to be too strict for B3LYP in \ac{SCF} method,
and so the calculations are not terminated for uracil, ibuprofen,
cholesterol, triadimenol, carbon dioxide and vitamin E,
which eventually leads to their divergence.
This again supports the overall robustness of the gradient descent scheme,
which terminates whenever the energy cannot be decreased more.
In other words,
the gradient descent always gives the monotone curves,
which is obviously not the case for \ac{SCF}.
Nevertheless, a considerable use of \ac{KAIN} with history of 5-6 previous
iterates fixes \ac{SCF} divergence issues.
In contrast,
acceleration of the multiwavelet version of the gradient descent method
is still under development.

\section{Grassmann manifold}
\label{Grassmann_manifold_section}
\setcounter{equation}{0}

The Grassmann manifold $\Gr(N)$
represents equivalence classes of orthonormal orbital sets
under unitary rotations.
Since the Hartree-Fock energy is invariant under such rotations,
optimization on $\Gr(N)$
eliminates redundant degrees of freedom.
We describe the induced Riemannian structure, characterize the horizontal space,
and derive the projected gradient and vector transport operators used
in Algorithm~\ref{algorithm_conjugate_gradient_descent}.
It is worth pointing out that while both the energy and 
$H^1$-norm are invariant with respect to the orbital rotations,
the $H^1$-inner product is obviously not.
Taking into account the importance of inner product for
the Riemannian optimization,
this lack of invariance gives rise to some complications that we treat
in this section.

The quotient $\Gr(N) = \St(N) / O(N)$
is defined by identifying the orbital vectors $\phi, \psi \in \St(N)$
that differ by some rotation into a single point $[\phi] \in \Gr(N)$.
The Stiefel and Grassmann manifolds are connected by means of the so called
canonical projection $\pi : \St(N) \to \Gr(N)$,
which is a smooth map defined by $\pi(\phi) = [\phi]$ with $\phi \in \St(N)$.
Its derivative $d \pi(\phi) : T_{\phi} \to T_{[\phi]}$ is a bounded linear operator
between the tangent spaces $T_{\phi} = T_{\phi} \St(N)$
to Stiefel at $\phi$
and $T_{[\phi]} = T_{[\phi]} \Gr(N)$ to Grassmann at $[\phi]$.
The former space $T_{\phi}$ was thoroughly described in the previous section.

Tangent vectors to $\Gr(N)$ are rather abstract objects.
Therefore, in practice elements of $T_{[\phi]}$ are represented by
particular elements of $T_{\phi}$.
For this purpose it is convenient to introduce a notion of fiber $F_{\phi}$
for $\phi \in \St(N)$,
that is the preimage $F_{\phi} = \pi^{-1} (\pi(\phi))$.
In other words, for a given orbital vector $\phi \in \St(N)$
the fiber coincides with the set
\[
    F_{\phi}
    =
    \{
        Q \phi
        :
        Q \in O(N)
    \}
    ,
\]
which is an embedded submanifold of $\St(N)$.
In particular,
the tangent space
\(
    V_{\phi} = T_{\phi} F_{\phi} 
\)
is a subspace of
\(
    T_{\phi} \St(N)
    .
\)
We refer to $V_{\phi}$ as the vertical space at $\phi$.
Its orthogonal complement $H_{\phi} = T_{\phi} \St(N) \ominus V_{\phi}$
is called the horizontal space at $\phi$.
It turns out that
\(
    V_{\phi} = \ker d \pi (\phi) 
\)
and
the restriction
\(
    d \pi (\phi) |_{H_{\phi}}
\)
of the derivative 
\(
    d \pi (\phi)
\)
to the horizontal space $H_{\phi}$
is bijective.
This linear isomorphism allows to represent every abstract tangent vector
at $[\phi]$ via a concrete horizontal vector at $\phi$,
that is obviously an element of the space $T_{\phi} = H_{\phi} \oplus V_{\phi}$
determined in the previous section.

From the optimization perspective the vertical component $V_{\phi}$
is the collection of all redundant directions
that are due to the rotational gauge invariance.
By projecting tangent vectors onto the horizontal subspace,
one eliminates these redundancies.
Importantly,
one can derive an explicit expression for this
horizontal projection,
denoted below by $\Proj_{\phi}^H$.
It is acting between $T_{\phi}$ and $H_{\phi}$.
Indeed, at first we characterize the vertical space $V_{\phi}$ as
the tangent to the fiber $F_{\phi}$ at $\phi$
in terms of smooth curves in $F_{\phi}$ passing through $\phi$.
Clearly, every such curve can be parametrized by $Q(t) \phi$,
where $Q(t)$ is a smooth function defined on an open interval containing zero
with the values $Q(t) \in O(N)$ and coinciding with the identity matrix at zero.
Thus
\[
    V_{\phi}
    =
    \left \{
        Q'(0) \phi
        :
        Q(t) \in O(N)
        \,
        \text{ is a smooth curve with }
        \,
        Q(0) = 1
    \right \}
    ,
\]
and therefore, we have
\begin{equation}
\label{vertical_space}
    V_{\phi}
    =
    \left \{
        \omega \phi
        :
        \omega
        \,
        \text{ is skew symmetric}
    \right \}
    .
\end{equation}
Indeed,
\(
    Q(t) = e^{\omega t}
\)
is an example of a smooth curve characterizing $V_{\phi}$
for an arbitrary skew symmetric matrix $\omega$,
which implies the inclusion
\(
    V_{\phi}
    \supset
    \text{RHS\eqref{vertical_space}}
    .
\)
By differentiating the equality
\(
    Q(t) Q(t)^T = 1
\)
at $t = 0$ and substituting $Q(0) = 1$,
one arrives at
\(
    Q(0) + Q'(0)^T = 0
\)
leading to the inverse inclusion
\(
    V_{\phi}
    \subset
    \text{RHS\eqref{vertical_space}}
    .
\)
Thus for $u \in T_{\phi}$ its horizontal projection
is $\Proj_{\phi}^H u = u - \omega \phi$,
where the skew symmetric matrix $\omega = \omega(u)$ can be obtained
from minimization of the following functional
\[
    \mathcal G(\omega)
    =
    \norm{u - \omega \phi}_{H^1}^2
    =
    \norm{u}_{H^1}^2
    -
    2
    \Innerprod{ u }{ \omega \phi}_{H^1}
    +
    \norm{\omega \phi}_{H^1}^2
    .
\]
Introducing two new matrices
$\Phi_{ij} = \Innerprod{\phi_i}{\phi_j}_{H^1}$
and $\Psi_{ij} = \Innerprod{u_i}{\phi_j}_{H^1}$
we can rewrite $\mathcal G(\omega)$ in terms of the Frobenius inner product
in the following form
\[
    \mathcal G(\omega)
    =
    \norm{u}_{H^1}^2
    -
    2
    \Innerprod{ \Psi }{ \omega }_F
    +
    \Innerprod{ \omega \Phi }{ \omega }_F
    .
\]
Then we compute its derivative $d \mathcal G(\omega)$
as
\[
    d \mathcal G(\omega) (\delta \omega)
    =
    -
    2
    \Innerprod{ \Psi }{ \delta \omega }_F
    +
    \Innerprod{ \omega \Phi }{ \delta \omega }_F
    +
    \Innerprod{ \delta \omega \Phi }{ \omega }_F
    =
    2
    \Innerprod{ \omega \Phi - \Psi }{ \delta \omega }_F
\]
for a skew symmetric matrix $\delta \omega$.
At the minimum we have $d \mathcal G(\omega) = 0$,
meaning that
\(
    d \mathcal G(\omega) (\delta \omega) = 0
\)
for all skew symmetric matrices $\delta \omega$,
and so one concludes that $\omega \Phi - \Psi$ is symmetric.
This leads to
\begin{equation}
\label{horizontal_sylvester_equation}
    \omega \Phi + \Phi \omega = \Psi - \Psi^T
    ,
\end{equation}
which uniquely determines the skew symmetric $N \times N$ matrix $\omega$.
Finding the solution of the Sylvester equation \eqref{horizontal_sylvester_equation}
reduces to diagonalizing $\Phi = U \Lambda_{\Phi} U^T$,
so that $\Lambda_{\Phi} = \diag (\lambda_1, \ldots, \lambda_N)$.
Thus
\[
    \omega = U Y U^T
    , \quad
    Y_{ij} = \frac{ S_{ij} }{ \lambda_i + \lambda_j }
    , \quad
    S = U^T \left( \Psi - \Psi^T \right) U
    .
\]
We conclude the description of the horizontal spaces
with the important formula
\begin{equation}
\label{horizontal_projection}
    \Proj_{\phi}^H u
    =
    u
    -
    \omega \phi
    , \quad
    \text{ for }
    \,
    u \in T_{\phi}
    .
\end{equation}

By means of the linear isomorphism
\(
    d \pi (\phi) |_{H_{\phi}}
    ,
    \phi \in \St(N)
    ,
\)
we can describe the geometry of the Grassmann manifold $\Gr(N)$.
For
\(
    \phi \in \St(N)
\)
and
\(
    \xi \in T_{[\phi]}
    =
    T_{[\phi]} \Gr(N)
\)
one defines the horizontal lift
by the formula
\(
    \lift_{\phi}(\xi)
    =
    \left(
        d \pi (\phi) |_{H_{\phi}}
    \right)^{-1}
    (\xi)
    ,
\)
which is a linear operator.
Composing the canonical projection derivative and the lift
one obtains
\[
    d \pi (\phi)
    \circ
    \lift_{\phi}
    =
    1_{ T_{[\phi]} }
    , \quad
    \lift_{\phi}
    \circ
    \,
    d \pi (\phi)
    =
    \Proj_{\phi}^H
    .
\]
Furthermore, we have the following identity
\begin{equation}
\label{lift_rotation}
    \lift_{ Q \phi }(\xi)
    =
    Q \lift_{ \phi }(\xi)
\end{equation}
for any
\(
    \phi \in \St(N)
    ,
    Q \in O(N)
\)
and
\(
    \xi \in T_{[\phi]}
    .
\)
Equation \eqref{lift_rotation} is proved in two steps.
Firstly, one will show that the right hand side of \eqref{lift_rotation}
is an element of the horizontal space at $\psi = Q \phi$.
For brevity we write
\(
    u = \lift_{ \phi }(\xi)
    ,
    v = \lift_{ \psi }(\xi)
    .
\)
The right hand side of \eqref{lift_rotation}
equals to $Qu$ and it belongs to the tangent space $T_{\psi}$
if and only if $Qu \in \ker d \mathcal Q(\psi)$
with $\mathcal Q$ defined by \eqref{definition_Q}.
The latter follows from
\[
    \Innerprod{(Qu)_i}{\psi_j}_{L^2}
    +
    \Innerprod{\psi_i}{(Qu)_j}_{L^2}
    =
    \sum_{k,l}
    Q_{ik} Q_{jl}
    \left(
        \Innerprod{u_k}{\phi_l}_{L^2}
        +
        \Innerprod{\phi_k}{u_l}_{L^2}
    \right)
    =
    0
\]
holding true for any $i, j$, because $u \in T_{\phi}$.
Now $Qu \in H_{\psi}$ follows from
\[
    \Innerprod{Qu}{\omega \psi}_{H^1}
    =
    \Innerprod{u}{ Q^T \omega Q \phi }_{H^1}
    =
    0
\]
holding true for any skew symmetric $\omega$,
since $Q^T \omega Q$ is skew symmetric and $u \in H_{\phi}$.
Thus both vectors $v, Qu \in H_{\psi}$
and in order to finish the proof of \eqref{lift_rotation},
that $v = Qu$, one needs to check that
\(
    d \pi(\psi)(v)
    =
    d \pi(\psi)(Qu)
    .
\)
The latter follows from the chain rule applied to the
corresponding curve representation.
Indeed,
if $c(t)$ is a smooth curve in $\St(N)$
passing through $\phi$ with the velocity $u$,
which means $c(0) = \phi$ and $c'(0) = u$,
then $\tilde c(t) = Q c(t)$ is a smooth curve in $\St(N)$
and $\bar c(t) = \pi(c(t))$ is a smooth curve in $\Gr(N)$
satisfying
\[
    \tilde c(0) = \psi = Q \phi
    , \quad
    \tilde c'(0) = Q u
    , \quad
    \bar c(0) = [ \phi ]
    , \quad
    \bar c'(0) = \xi
    .
\]
Therefore, since $\pi \circ \tilde c(t)$ is a smooth curve
passing through $[\psi]$ we have
\[
    d \pi(\psi)(Qu)
    =
    ( \pi \circ \tilde c )'(0)
    =
    ( \pi \circ c )'(0)
    =
    \bar c'(0)
    =
    \xi
    =
    d \pi(\psi) \left( \lift_{\psi} \xi \right)
    =
    d \pi(\psi)(v)
    .
\]
This concludes the proof of \eqref{lift_rotation}.

Equation \eqref{lift_rotation} allows us to define an inner product
in $T_{[\phi]}$ unambiguously,
turning $\Gr(N)$ into a Riemannian manifold.
Abusing the $H^1$ notation we define
\begin{equation*}
    \Innerprod{\xi}{\zeta}_{H^1}
    =
    \Innerprod{ \lift_{\phi} \xi }{ \lift_{\phi} \zeta }_{H^1}
\end{equation*}
for $\xi, \zeta \in T_{[\phi]}$,
where $\phi$ is a representative of the equivalence class $[\phi]$.
Obviously, this product does not depend on
the choice of representative by \eqref{lift_rotation}.
Furthermore, it is straightforward to show that the lift at $\phi \in \St(N)$ of
the Riemannian gradient defined with respect to Grassmann manifold
coincides with the Riemannian gradient over the Stiefel manifold
\begin{equation}
\label{grassmann_energy_gradient}
    \lift_{\phi}
    \grad \mathcal E ([\phi])
    =
    \grad \mathcal E \circ \pi (\phi)
    ,
\end{equation}
where $\mathcal E$ is regarded as a functional defined on $\Gr(N)$.
The right hand side of \eqref{grassmann_energy_gradient}
was determined in the previous section,
see Equation \eqref{grad_E}.

\begin{figure}[ht!]
\centering
\includegraphics[width=0.49\textwidth]
{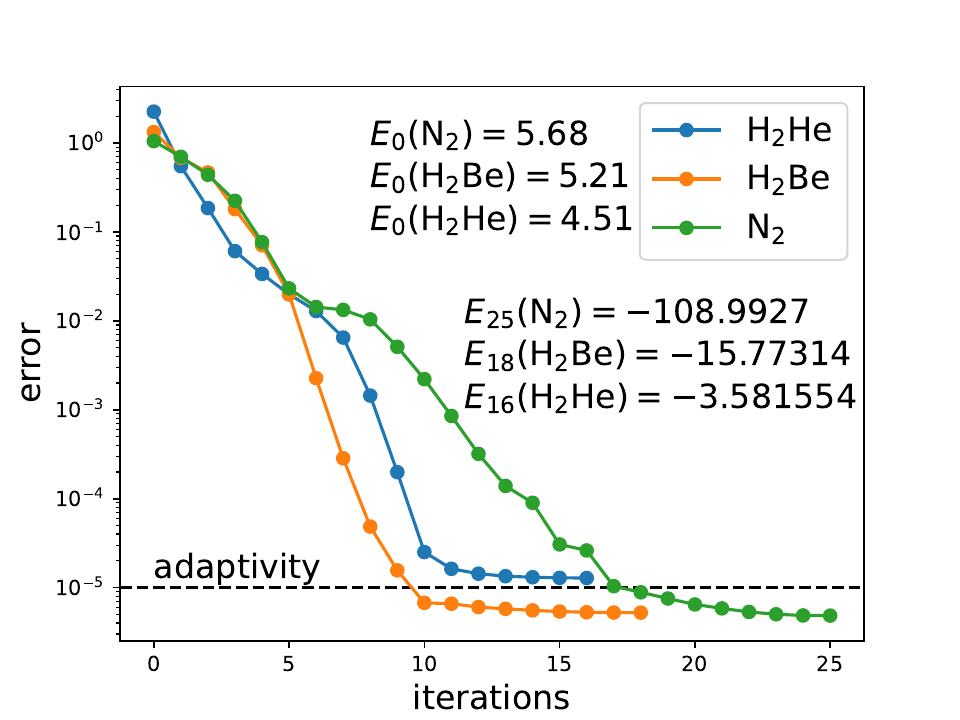}
\includegraphics[width=0.49\textwidth]
{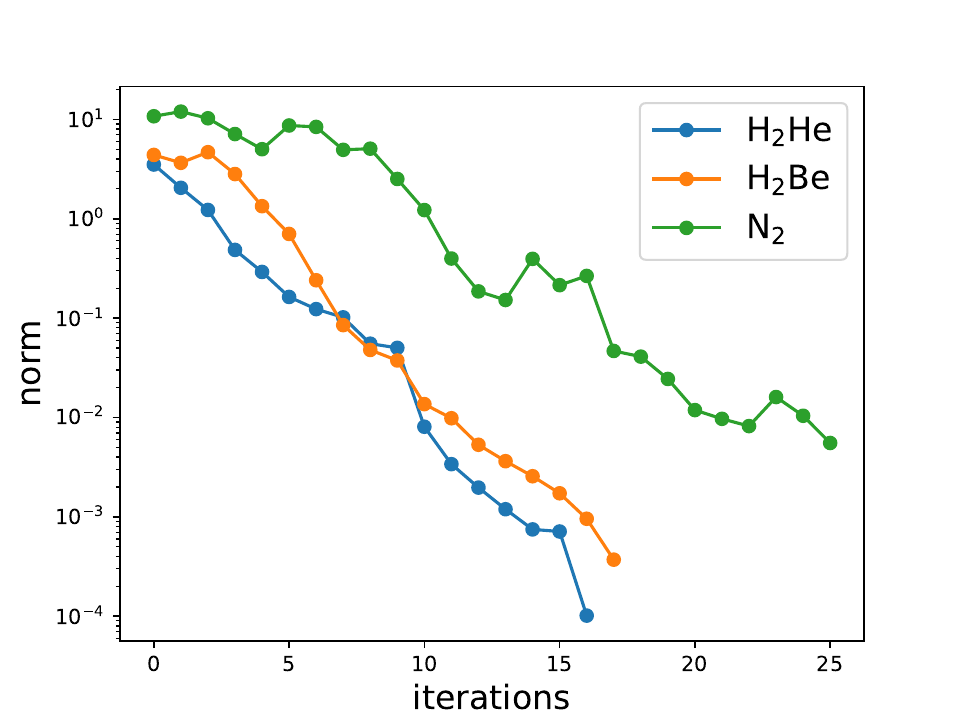}
\caption{
    Convergence of the Riemannian conjugate gradient descent
    on the Grassmann manifold for
    Hartree-Fock model starting from a random Gaussian superposition.
}
\label{fig:grassmann_random_convergence}
\end{figure}

We introduce a retraction as the following equivalence class
\begin{equation}
\label{grassmann_retraction}
    \mathcal R_{[\phi]} (\xi)
    =
    [ \mathcal R_{\phi}( \lift_{\phi} \xi ) ]
    , \quad
    \xi \in T_{[\phi]}
    ,
\end{equation}
where $\mathcal R_{\phi}$ is defined in \eqref{lowdin_retraction}.
Let us show that this definition is well defined, namely,
it does not depend on the choice of a representative $\phi \in \St(N)$
in the class $[\phi] \in \Gr(N)$.
Indeed,
let $Q \in O(N)$.
We need to show that $\text{RHS}(\ref{grassmann_retraction})$
will not change after substituting $Q \phi$ in place of $\phi$.
The orbital vector $\phi + \lift_{\phi} \xi$ has the following overlap matrix
\[
    S(\phi + \lift_{\phi} \xi)
    =
    C(\phi, \phi)
    +
    C(\phi, \lift_{\phi} \xi)
    +
    C(\lift_{\phi} \xi, \phi)
    +
    C(\lift_{\phi} \xi, \lift_{\phi} \xi)
    =
    1 + S(\lift_{\phi} \xi)
    .
\]
Notice
\(
    C(Qu, Qv) = Q C(u, v) Q^T
\)
for any orbital vectors $u, v$.
Then using \eqref{lift_rotation} one obtains
\[
    S(Q\phi + \lift_{Q\phi} \xi)
    =
    S(Q(\phi + \lift_{\phi} \xi))
    =
    1 + Q S(\lift_{\phi} \xi) Q^T
    .
\]
If $U \in O(N)$ diagonalizes $S(\lift_{\phi} \xi)$
then $QU$ diagonalizes $Q S(\lift_{\phi} \xi) Q^T$.
Hence
\[
    \mathcal R_{Q\phi}( \lift_{Q\phi} \xi )
    =
    QU (1 + \Lambda)^{-1/2}(QU)^T
    (Q\phi + \lift_{Q\phi} \xi)
    =
    Q
    \mathcal R_{\phi}( \lift_{\phi} \xi )
    ,
\]
which justifies the definition \eqref{grassmann_retraction}.

\begin{figure}[ht!]
\centering
\includegraphics[width=0.49\textwidth]
{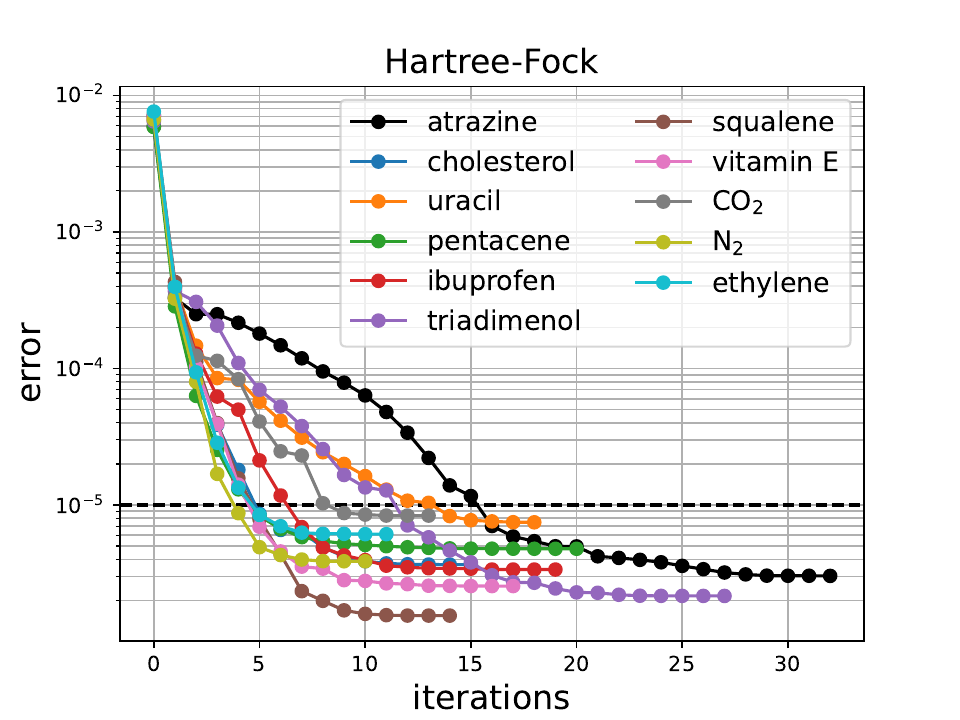}
\includegraphics[width=0.49\textwidth]
{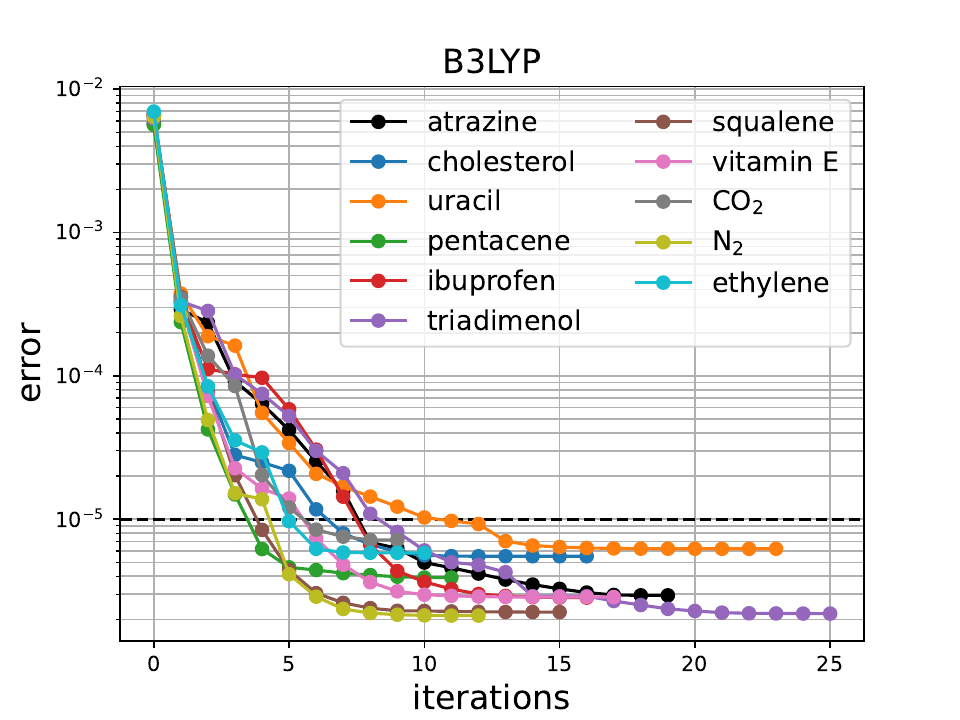}
\caption{
    Convergence of the Riemannian conjugate gradient descent
    on the Grassmann manifold for
    Hartree-Fock and B3LYP energy models.
}
\label{fig:grassmann_mrchem_energy_convergence}
\end{figure}

By working with the horizontal vectors in $H_{\phi}$
instead of the tangent vectors in $T_{[\phi]}$
and accounting for \eqref{grassmann_energy_gradient}, \eqref{grassmann_retraction},
one can see that the steepest gradient descent algorithm
\eqref{steepest_descent} does not change by moving to the Grassmann manifold.
In other words,  the steepest descent is not affected by the gauge invariance.
On the contrary,
the conjugate descent algorithm \ref{algorithm_conjugate_gradient_descent}
is affected by this invariance through preconditioning and vector transporting.
A preconditioner is introduced as the following composition
\begin{equation*}
    \Prec_{\phi}
    =
    \Proj_{\phi}^H \Proj_{\phi} T^{-1}
\end{equation*}
acting in the horizontal space $H_{\phi}$.
This unambiguously determines a preconditioner in the tangent space $T_{[\phi]}$.
We can also introduce a transporter by lifting
the transporter
\(
    \mathcal T_{ \phi \leftarrow \psi }
    =
    \Proj_{\phi}
\)
used on the Stiefel manifold
and projecting to the horizontal space $H_{\phi}$.
In other words,
the restriction
\(
    \Proj_{\phi}^H \Proj_{\phi} |_{H_{\psi}}
\)
through the lifting procedure defines
a transporter
\(
    \mathcal T_{ [\phi] \leftarrow [\psi] }
\)
on the Grassmann manifold.
This concludes the description of Algorithm \ref{algorithm_conjugate_gradient_descent}
in the case of Grassmann manifold.
Notably, in practice the Stiefel calculations
are modified by essentially adding an additional projection.
Thus we can repeat the calculations from the previous section,
presented in Figures \ref{fig:stiefel_random_convergence}
and \ref{fig:stiefel_mrchem_energy_convergence},
while eliminating the rotational degrees of freedom.
The corresponding results are given in Figures \ref{fig:grassmann_random_convergence}
and \ref{fig:grassmann_mrchem_energy_convergence}, respectively.

\vskip 0.05in
\noindent
{\bf Acknowledgments.}
{
    The author acknowledges support from the Research Council of Norway
    through its Centres of Excellence scheme (Hylleraas centre, 262695),
    and from NOTUR -- The Norwegian Metacenter for Computational Science
    through grant of computer time (nn14654k).
}

\bibliographystyle{acm}
\bibliography{bibliography}

\end{document}